\documentclass[a4paper,fleqn,usenatbib]{mnras}
\usepackage{newtxtext,newtxmath}

\usepackage[T1]{fontenc}
\usepackage{ae,aecompl}

\usepackage{graphicx}	
\usepackage{amsmath}	

\graphicspath{{figures/}}
\usepackage{enumitem}

\title[Fragmentation of a large space body]{Effect of the surface shape of a large space body on its fragmentation in a planetary atmosphere}

\author[D.E. Khrennikov et al.]{
Daniil E. Khrennikov,$^{1}$
Andrei K. Titov,$^{2}$
Alexander E. Ershov,$^{1,3}$
\newauthor Andrei B. Klyuchantsev,$^{4}$
Vladimir I. Pariev$^{5}$\thanks{E-mail: vpariev@td.lpi.ru}
and Sergei V. Karpov$^{1,4,6}$\thanks{E-mail: karpov@iph.krasn.ru}
\\
$^{1}$Siberian Federal University,   Svobodny Av. 79/10, Krasnoayrsk,  660041, Russia\\
$^{2}$Moscow Institute of Physics and Technology, Institusky Per. 9, Dolgoprudny 141700, Russia\\
$^{3}$Institute of Computational Modeling SB RAS,  Akademgorodok 50/44, Krasnoyarsk, 660036, Russia\\
$^{4}$L.~V. Kirensky Institute of Physics, Federal Research Center KSC SB RAS,  Akademgorodok 50/38, Krasnoyarsk, 660036, Russia \\
$^{5}$P.~N. Lebedev Physical Institute, Leninsky Prosp. 53, Moscow 119991, Russia \\
$^{6}$Siberian State University of Science and Technology, Krasnoyarsky Rabochy Av. 31, Krasnoyarsk 660014, Russia
}

\date{Accepted 2020 January 27. Received 2020 January 25; in original form 2019 June 28}

\pubyear{2020}

\begin{document}
\label{firstpage}
\pagerange{\pageref{firstpage}--\pageref{lastpage}}
\maketitle

\begin{abstract}
Employing the finite element and  computational fluid dynamics methods, we have determined the conditions for the fragmentation of space bodies or preservation of their integrity when they penetrate into the Earth's atmosphere. The origin of forces contributing to the fragmentation of space iron bodies during the passage through the dense layers of the planetary atmosphere has been studied. It was shown that the irregular shape of the surface can produce transverse aerodynamic forces capable of causing deformation stress in the body exceeding the tensile strength threshold of iron. 
\end{abstract}




\section{Introduction}

One of the important problems in the modern astronomy is forecasting the collision of the Earth with asteroids, capable of causing extensive damage to the planet. In this context the study of the passage of large space bodies (SB) through the planetary atmosphere is of great interest~\citep{Stulov1995,Bronshten1981,Martin1966,Loh1963,Hawkins1964,Adushkin1994,Morrison1994,Cotto-Figueroa2016,Khrennikov2019}. Such processes are accompanied by the fragmentation of space bodies, which reduces the degree of damage to terrestrial objects due to the fall onto the planet surface of individual fragments with significantly less mass and kinetic energy. Thus, one of the important aspects of the problem of space safety of the Earth is the prediction of the possibility of destruction of an SB and scattering of its fragments in the atmosphere~\citep{Nemchinov1999,Cotto-Figueroa2016}.

Among the factors determining the probability of fragmentation are the irregular shape of an SB, which dramatically deviates from a sphere, SB material, the velocity of entry into the atmosphere, the length of the trajectory, the density of the atmosphere, and a number of other factors. Forecasting the fragmentation and the degree of danger of large SBs can be potentially based on the advance information on the shape and type of the material of SBs obtained from radio astronomical observations \citep{Thompson2017,Naidu2015,Greenberg2017}.

At present there are a number of models of fragmentation considered in publications: e.g.,~\citet{Nemchinov1999,Svetsov1995} provide the overviews of existing concepts on the fragmentation processes of large SBs. In particular, within the framework of the dynamic model of fragmentation, the motion of fragments is considered as the motion of individual bodies diverging due to aerodynamic interaction. In addition, the influence of aerodynamic pressure on internal cracks of an SB is very important, which can cause subsequent destruction of large fragments. Another fragmentation model called a ``sandbag'' is also discussed, which represents the SB in the form of a cluster of fragments of different sizes moving in one cloud~\citep{TeterevNemchinov93}. For large SBs, the fragmentation process is considered using the model of the liquid drop, which is strongly deformed in the area of maximum pressure at the surface \citep{Adushkin1994, Svetsov1995}. The formation of a concave surface in this area is considered as the initial stage of disintegration of SB into two parts \citep{Baldwin1971, Svetsov1995}.

A number of computational and theoretical models for the fragmentation phase had been developed using various physical approaches \citep[ e.g., see][]{Nemchinov1999,Ivanov2005,Fadeenko1997,Grigoryan1979,Hills1993,Stulov1998,Svetsov1995,Barri2010}. General feature of the models is that they use the destruction criteria adopted in the mechanics of materials, although these criteria are considered to be unacceptable for breaking brittle bodies (e.~g., structurally heterogeneous SBs). As a result, fragmentation is considered as a total destruction~\citep{Nemchinov1999} producing very large number of tiny fragments, or as successive ``avalanche'' ruptures, in which the body disintegrates into smaller pieces~\citep{Nemchinov1999,Stulov1998,Svetsov1995,Weeler2018}. In papers~\citet{Ivanov1999a,Ivanov1999b} the model for sequential fragmentation of SBs was presented, first, the main body fragments, and then its fragments disintegrate in turn. The model is based on the existence of the tensile strength threshold value and modern fracture mechanics. 
The discrete fragmentation model was used to determine the average fragment size at each fragmentation step, as well as the coordinates of the trajectory points where this process occurs. In addition, it was asserted in~\citet{Ivanov1999a} that for small sized SBs their shape (spherical and cuboid) does not play a role in fragmentation and calculations of their interaction with the atmosphere. A review of some models of destruction and scattering of SBs composed of robust materials in the atmosphere is given, e.~g. in~\citet{Artemeva1994}. On the basis of these models, the authors of the  paper~\citet{Artemeva1994} performed numerical 3D calculations of the scattering of fragments of two identical SBs in the atmosphere. These fragments were initially placed in close proximity to one another, and were pushed apart by the aerodynamic pressure forces. 

It should be noted that most of  these models do not pay  attention to such an important factor as the  shape of an SB, containing concavities and recesses. The shapes of large SBs can be divided into regular (close to the sphere) and irregular with the chaotic topography of the surface. When this factor is taken into account, the conditions can arise under which the aerodynamic forces different from those on a sphere are capable to produce  deformation and rupture of the SB.

The material of an SB is another important characteristic that determines the resistance of the SB to fragmentation. Inhomogeneous stone SBs are characterized by the presence of deep internal microcracks, which on the frontal surface of the SB are filled with atmospheric gases of the shock wave boundary layer with ultrahigh pressure. These cracks promote the appearance of transverse forces and the fracture of the SB.

The tensile strength of stone SBs is low (1-5 MPa)  \citep{Petrovic2001,Pisarenko1975}, primarily because of their intrinsic heterogeneity. Iron and iron-nickel SBs are homogeneous with no internal cracks, which significantly increases their tensile strength  (on average to 170-400 MPa and higher) compared to stone  \citep{Kaye1941,Pisarenko1975,Petrovic2001,Cotto-Figueroa2016}. Herewith, compressive strength of stone (granite) (120-200 MPa)  \citep{Pisarenko1975,Petrovic2001} is close to the minimum value of tensile strength of iron.  Despite the polycrystalline structure, the iron SBs are the  most resistant to fragmentation (the minimum compressive strength is 430 MPa \citep{Petrovic2001}) especially with increasing nickel fraction and a decrease in temperature \citep{Petrovic2001}. Note, that materials of meteorites exhibit considerable scatter in their strengths as well as traditional terrestrial materials do even with a slight difference in composition.
Moreover the same terrestrial and SB materials can exhibit a difference in strength \citep{Cotto-Figueroa2016}. 
We emphasize that the influence of the surface shape in the form of concavities and recesses of large SBs on the ability to fragment was not discussed in the literature.

The goal of our paper is to study the effect of the shape of large non-spherical iron SBs on their susceptibility to fragmentation due to the aerodynamic pressure on the protrusions and concavities at the surface, as well as to determine the origin of transverse forces that act on large surface defects and are able to rupture SB. Within the framework of this paper, the transverse forces are evaluated for a structurally homogeneous large SB comprised of iron as the strongest material and are compared to the threshold value of tensile stress, above which the fragmentation of the SB is triggered. We also determine the conditions for preserving the integrity of the large iron SB with radius 50---100 m during its through passage of the atmosphere at a minimum trajectory altitude of 10-15 km.

Note that examples of large iron SBs are well known: the Arizona meteorite with an evaluated initial size over 50~m, and the Hoba meteorite with the size of the largest found fragment about 3~m. 

\section{The model of space body and computational method }

\begin{figure}
\centering
	\begin{tabular}{cc}
	\includegraphics[width=2.5cm]{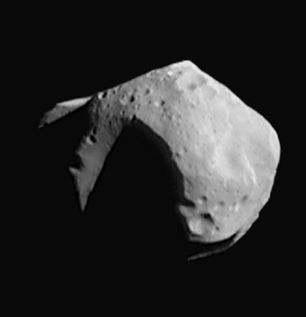}&
	\includegraphics[width=3cm]{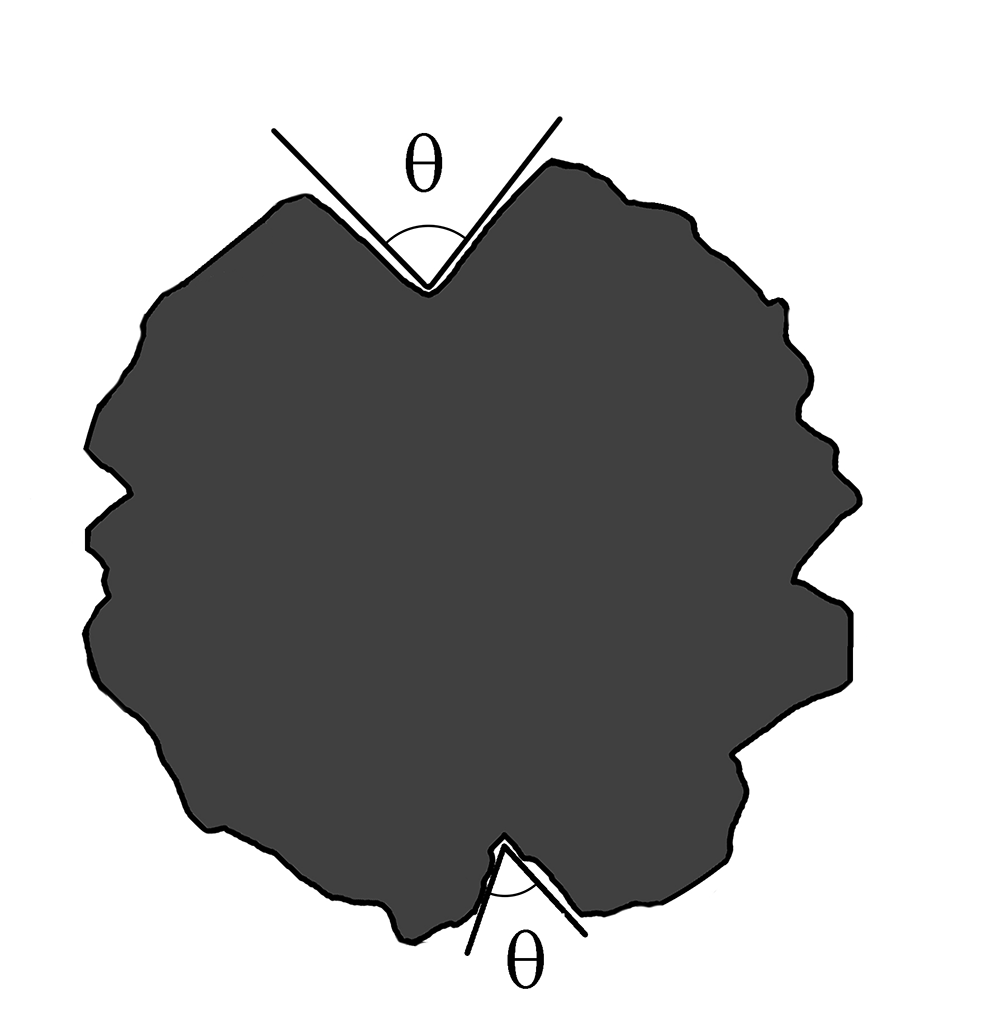}\\
	(a)&(b)\\
	\includegraphics[width=2.5cm]{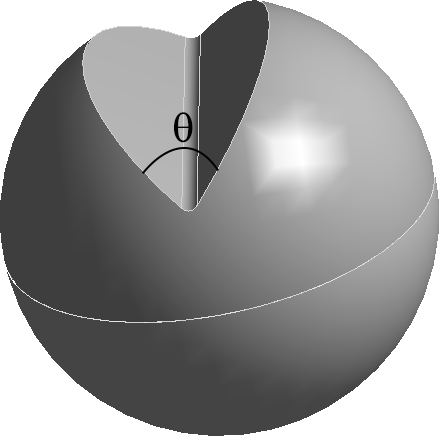}&
	\includegraphics[width=3cm]{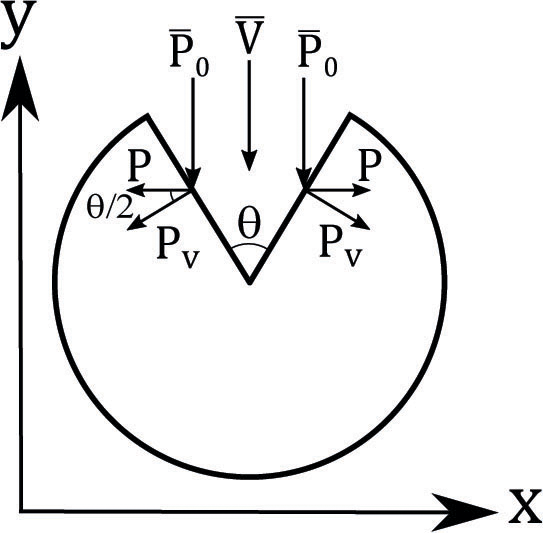}\\
	(c)&(d)
	\end{tabular}
    \caption{The irregular shape of the SB profile (a, b), the idealized model shape of the surface defect (c) and the distribution of the aerodynamic pressure on the flat sides of the defect (d), which can cause the rupture of SB (here $\theta$ is the opening angle).}
    \label{fig1}
\end{figure}

First of all, we have  to explain the origin of the aerodynamic force acting both on stretching and on compression of SB for such shapes as  cone, sphere, sphere with a defect shaped as an extended notch. The characteristic shapes of known asteroids and the comet nuclei of sufficiently large size (of the order of several kilometers) are irregular (Fig.~\ref{fig1}a). SBs of smaller size, as a rule, are characterized by an even greater deviation of the surface from the sphere. 

Fig.~\ref{fig1}b, as an example, schematically shows the section of the SB profile with numerous  defects in the shape of a triangular recess. An idealized model shape of such a defect could be a cutout in the shape of a spherical wedge, shown in Fig.~\ref{fig1}c. In this case, the SB has flat sides of the notch. 
Of course, ordinary craters are the prevalent  type of the SB surface defects. However, they are characteristic of larger bodies. Gravity of the SBs with sizes  considered in our work is too low, but this factor is important in formation of a classical crater with quasi-spherical bottom shape. In our case of  moderate size SBs, more likely result of a collision of the SB with a meteoroid is a conical notch~\citep{Melosh1989,Giese2006,Giese2014}.
We start formulating the problem keeping in mind a wedge-shaped notch such as shown in Fig.~\ref{fig1}c, and then also perform calculations for a conical notch such as shown in the bottom panel in Fig.~\ref{fig5}. Similar configurations of different lengths and depths may be located equally likely at different places of the surface.

The directions of the pressure forces on sides of the notch are shown in Fig.~\ref{fig1}d. The projections of these forces, perpendicular to the air flow velocity vector ($V$), create transverse deformation of the body, which can cause the rupture of SB. The wedge is characterized by two parameters: the opening angle $\theta$ and the depth of the wedge $h$ from the surface of the sphere, which approximates the shape of the SB.  
Within this reasoning, the forms most resistant to fragmentation would be a cone and, to lesser extent, a sphere. This is due to that the aerodynamic lateral forces act on the transverse compression of cones 
and spheres. 

Calculations of deformation and stress in SB are carried out using  the software  package ANSYS~\citep{ansys} for static equilibrium. ANSYS  is based on  the equations of elasticity derived, for example, in~\citet{Landau7}. For each volume element of a deformed body, in case of  equilibrium of forces ($\sum F_i=0$) the condition 
 is satisfied
\begin{equation}
    \sum_{k=1}^3 \frac{\partial\sigma_{ik}}{\partial x_k}=0,
\end{equation}
where $\sigma_{ik}$ is the stress tensor, equal to $i$-th component of the force, acting per unit surface, perpendicular to the axis $k$. 

Each index in equations $i$, $k$, $l$ takes values $1$, $2$, $3$, and corresponds to coordinate axes.

The conditions of equilibrium at the boundaries of the body are described by the equation
\begin{equation}
    P_i=\sum_{k=1}^3 \sigma_{ik}n_k,
\end{equation}
where $n_k$ is the normal vector to the surface of the body, $P_i$ is the vector of pressure and external forces.

The stress tensor is related to the strain tensor by the following equation \citep{Landau7}
\begin{equation}
    \sigma_{ik}=K\sum_{l=1}^3 u_{ll} \delta_{ik}+2\mu \left(u_{ik}-\frac{1}{3}\delta_{ik}\sum_{l=1}^3 u_{ll}\right).
\end{equation}

Here $K$ and $\mu$ are the bulk compression modulus and the shear modulus of a given material, respectively, $\delta_{ik}$ is Kronecker's symbol, $u_{ik}$ is the strain tensor, which is given by the equation \citep{Landau7}
\begin{equation}
    u_{ik}=\frac{1}{2}\left( \frac{\partial u_i}{\partial x_k}+\frac{\partial u_k}{\partial x_i}+ \sum_{l=1}^3 \frac{\partial u_l}{\partial x_i} \frac{\partial u_l}{\partial x_k}\right),
\end{equation}
where $u_i$ is the deformation vector, defined as
\begin{equation*}
    u_i=x_i' -x_i,
\end{equation*}
$x_i'$, $x_i$ are the components of the position vectors ${\bf x}'$ and $\bf x$ after and before deformation.

Besides that, we have to  assess the effect of heating the material of the SB and changes of its elastic properties due to heat transfer to the inner layers. When SB moves in the dense layers of the atmosphere, its frontal surface is heated and this heat is redistributed into the deeper layers of the SB.

To estimate the temperature distribution inside the SB and possible changes in the elastic properties of the material, we have to solve the 1D equation of thermal conductivity for the point of the SB surface with maximum temperature 

\begin{equation}
    \frac{\partial T}{\partial t} - \chi \frac{\partial^2 T}{\partial x^2}=0.
    \label{TD}
\end{equation}
Here $x$ is the coordinate along which a heat propagates, $\chi=\frac{\varkappa}{\rho c_p}$ is the thermal diffusivity of the SB material, $\rho$ is its density, $c_p$ is isobaric specific heat, $\varkappa$ is the coefficient of thermal conductivity. 

The initial and boundary conditions of the problem are given in the form:
\begin{equation*}
    T(x=0)=T_s,
\end{equation*}
where $T_s$ is the initial temperature at the SB surface assumed to be $10000\,\mbox{K}$ (see 
properties of strong radiative shocks in, e.g., \citet{Zeldovish1967}).

\section{Results and discussion}
\begin{figure}
\centering
	\includegraphics[width=5cm]
	{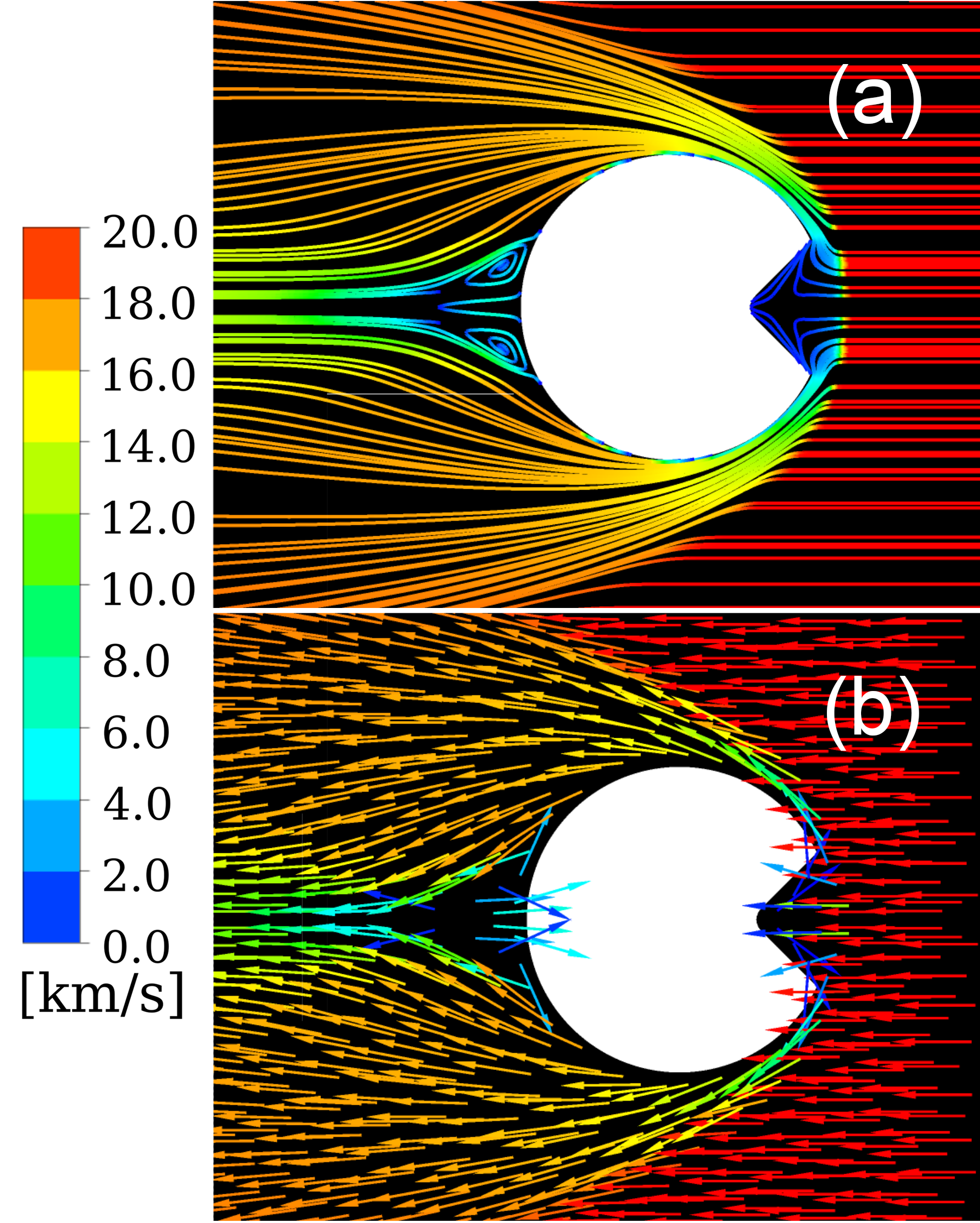}
    \caption{Air velocity around iron SB with radius $R=100$~m; (a)~--- streamline configuration, (b)~--- velocity vector field.}
    \label{fig2}
\end{figure}

Fig.~\ref{fig2} shows the pattern of distribution of the velocity field  around SB at the velocity $V=20$~km/s. The results were obtained with the software package ANSYS Fluent~\citep{ansys} for computational fluid dynamics. This package is a universal software system of the finite volume method applied for solving various problems in aero- and hydrodynamics. To calculate pressure distribution at the surface of SB with the effect of air compression due to pressure-density dependence the ANSYS Fluent ``ideal gas'' model was used with the laminar flow regime, hypersonic flow regime, density-based solver, and pressure-far-field option for inlet boundary to access high Mach number conditions.   Ambient static pressure was set to $26500\,\mbox{Pa}$.  The altitude of SB above the Earth surface is  10 km. The default air temperature was set to 219~K~\citep{ICAO}. 

The structures of aerodynamic fluxes around a
spherical SB as well as  around a spherical SB with wedge-shaped notch with  formation of vortex structures  and stagnation zones in 
its rear and front parts are shown in our paper \citep{Khrennikov2019} and in Fig.~\ref{fig2}.  
ANSYS Fluent makes it possible to calculate an accurate distribution of the static pressure on the surface of SB with arbitrary shape by default. 
In turn, this allows one to calculate accurately the distribution of a strain  stress in such a body by ANSYS Mechanical~\citep{ansys}.

\begin{figure}
\begin{tabular}{c}
	\includegraphics[width=8cm]{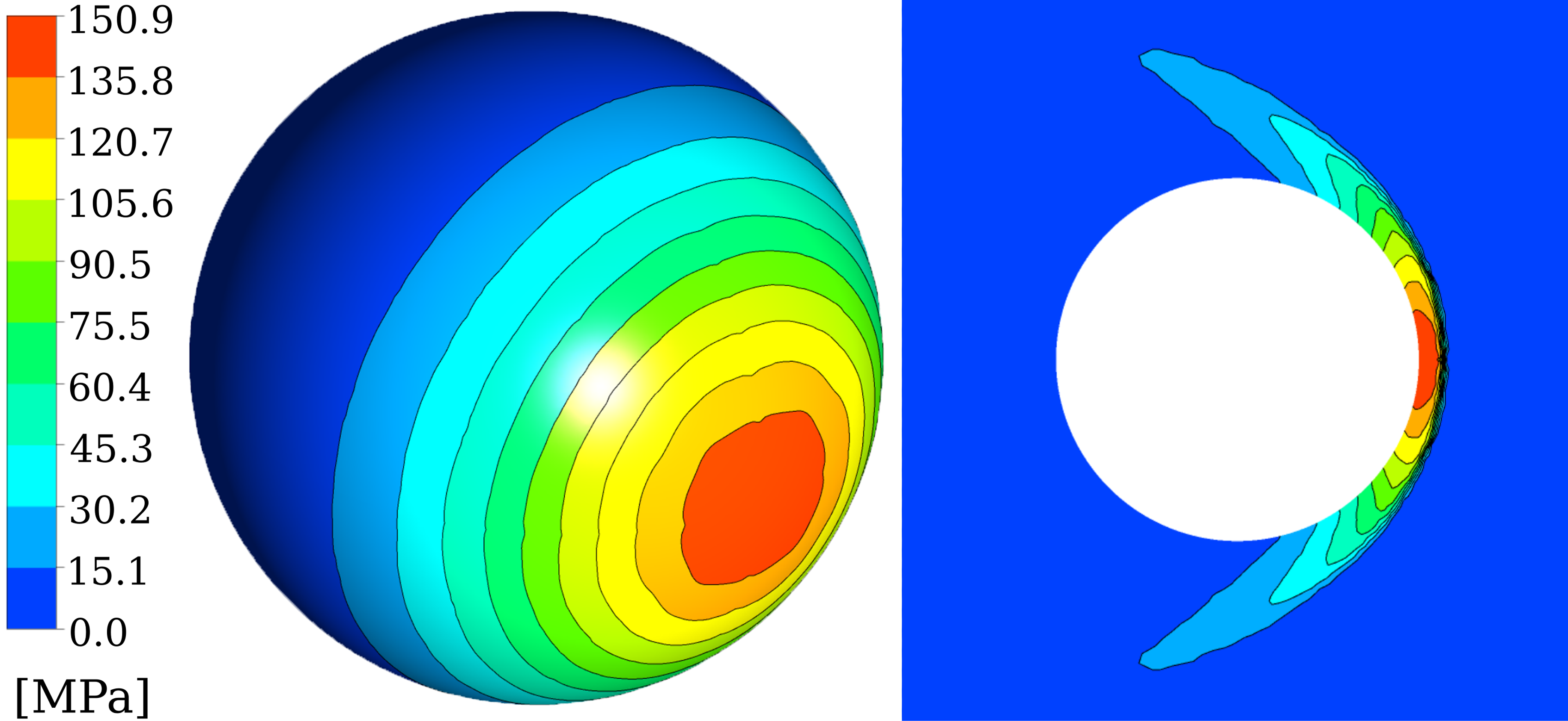}\\
	\includegraphics[width=8cm]{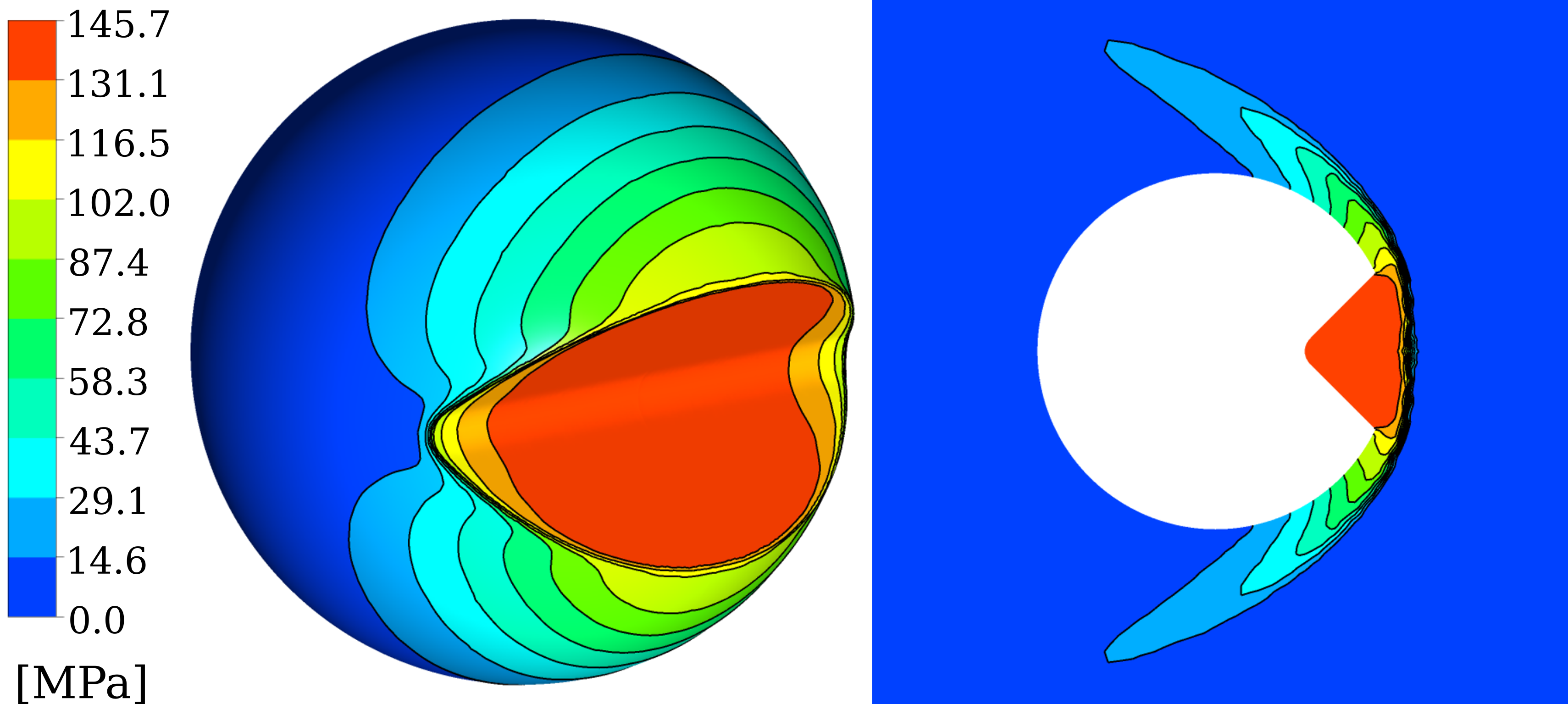}
	 \end{tabular}
    \caption{Comparative distributions of static pressure  at the iron SB surface (left),  and around SB (right) for two shapes: sphere (on top) and sphere with wedge (at the bottom). $R=100\,\mbox{m}$, $V=20\,\mbox{km/s}$,  the trajectory altitude is 10 km.}
    \label{fig3}
\end{figure}

Fig.~\ref{fig3} demonstrates the distribution of static pressure at the surface as well as around spherical SB with the radius $R=100\,\mbox{m}$ and  spherical SB with the wedge-shaped notch  shown in Fig.~\ref{fig1}c. The parameters of the latter SB are: the radius is $R=100\,\mbox{m}$, the curvature radius of the notch bottom $r_0=5\,\mbox{m}$, the opening angle of the notch is $90^\circ$, the depth of the notch is $h=0.5R$. Computations show that in the second case the maximum pressure is concentrated at the flat faces of a notch.

The simplified Eq.~\eqref{eq2} serves only for approximate estimates of external aerodynamic pressure on the faces of a wedge-shaped notch on the SB and for interpretation of the origin of transverse forces acting on large surface defects and 
the effect of the wedge opening angle  (Fig.~\ref{fig3}). Numerical simulations with ANSYS provide more accuracy, 
especially, in the vicinity of sharp edges.

\begin{figure}
\centering
	\includegraphics[width=5cm]{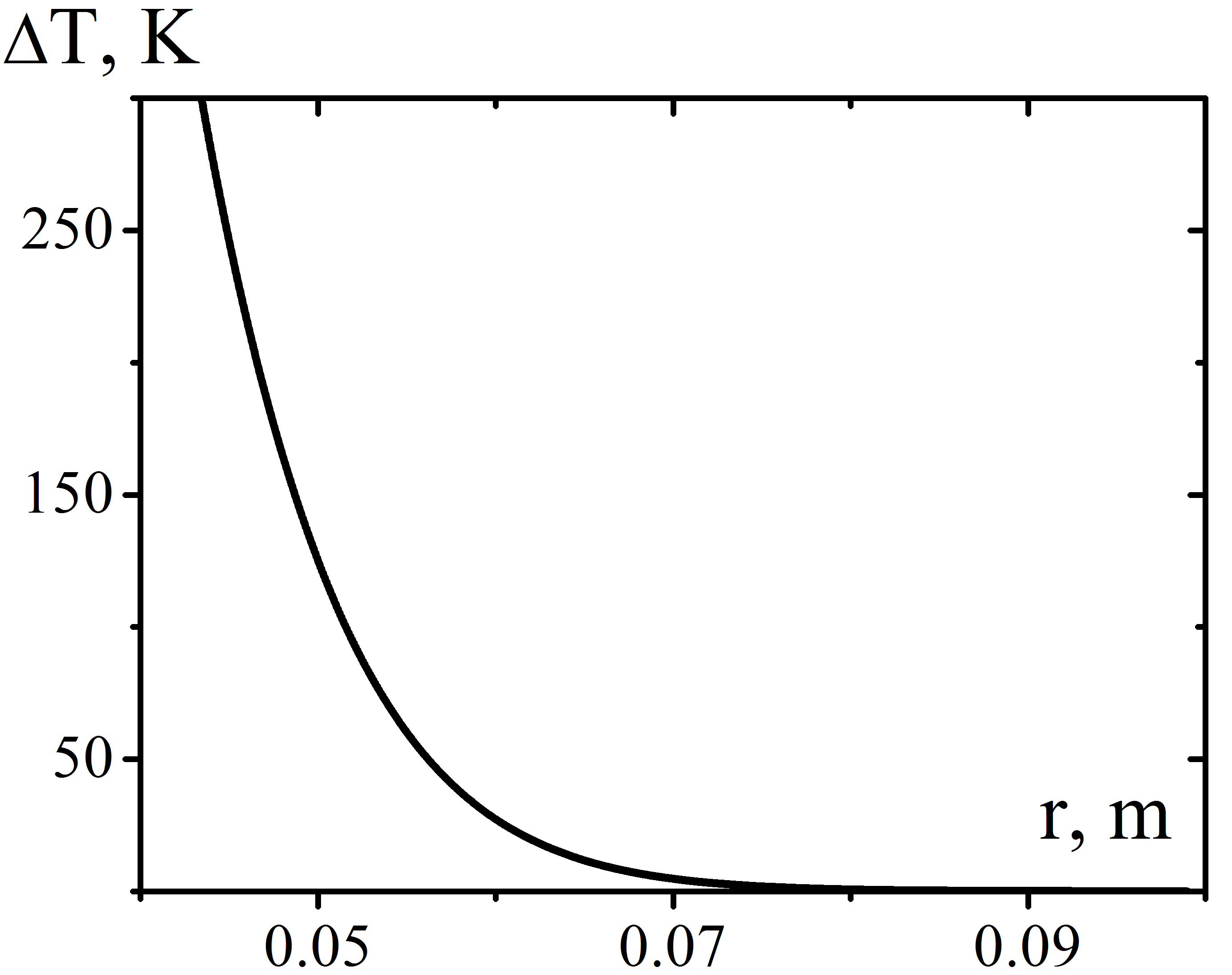}
	    \caption{Temperature distribution with the depth from the iron SB surface. The surface temperature is assumed to be $10^4$~K, the heating time is 10~s. The thermal diffusivity of the SB material (iron) is $2.3\cdot 10^{-5}$m$^2$/s~\protect\citep{Monaghan2001}.}
    \label{fig4}
\end{figure}

Fig.~\ref{fig4} shows the temperature variation from the frontal  surface of a large SB in the direction to its centre due to thermal conductivity. $\Delta T$ is the difference between the initial SB temperature  taken to be $219$~K throughout the depth, and the temperature profile after heating of the surface for 10 seconds and  fixed surface temperature 10000~K. These results were obtained by solving the 1D heat conduction equation~\eqref{TD}. As follows from the figure, the inner, deeper layers of SB are heated insignificantly during the passage of SB through dense layers of the atmosphere  and the heating layer is limited to a depth of about 7~cm. This leads us to the conclusion that such heating does not significantly  affect the elasticity modulus of the material within  the entire SB and its strength. 

\begin{figure}
\centering
    \begin{tabular}{c}
	\includegraphics[width=6cm]{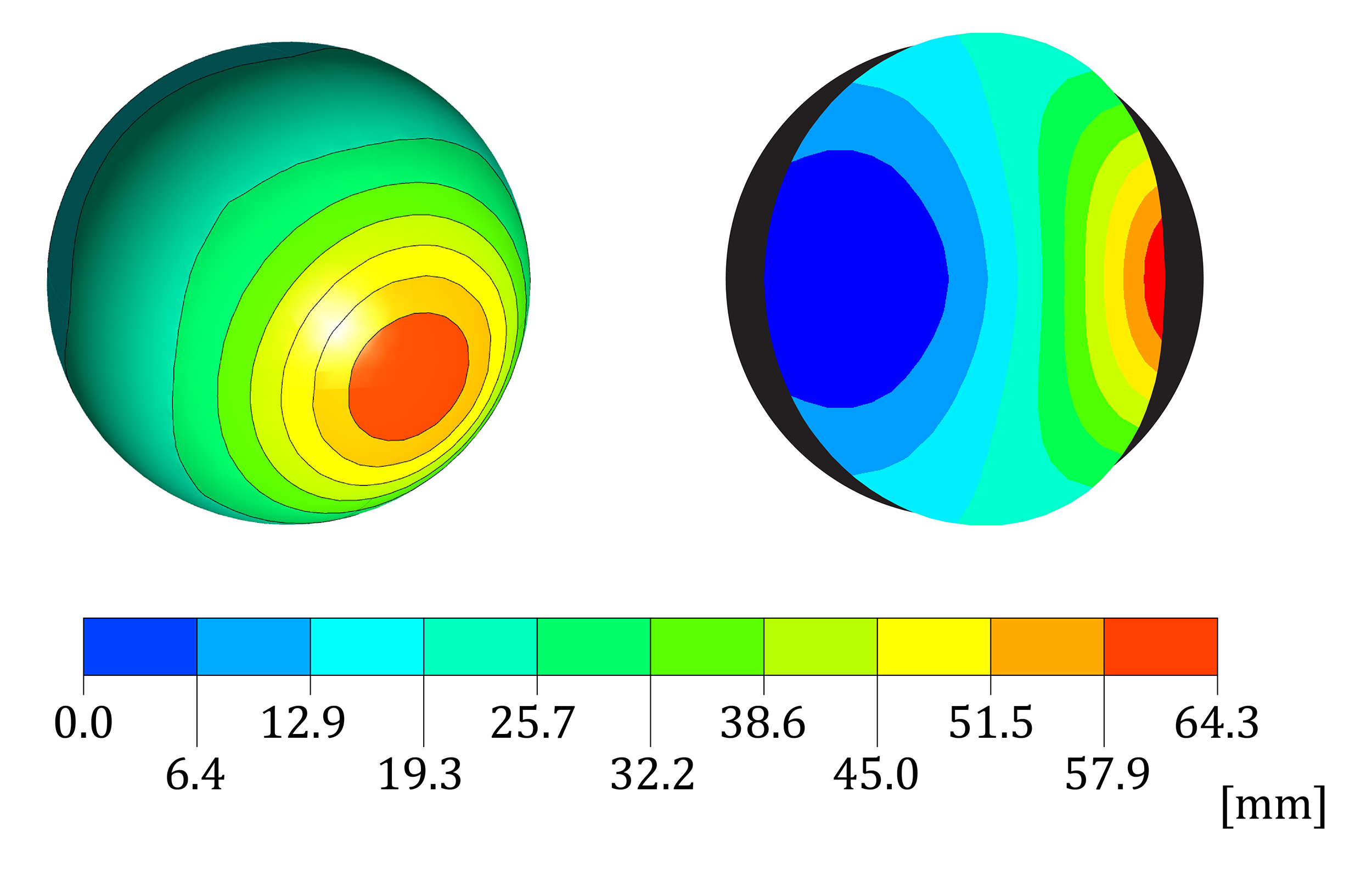}\\
	\includegraphics[width=6cm]{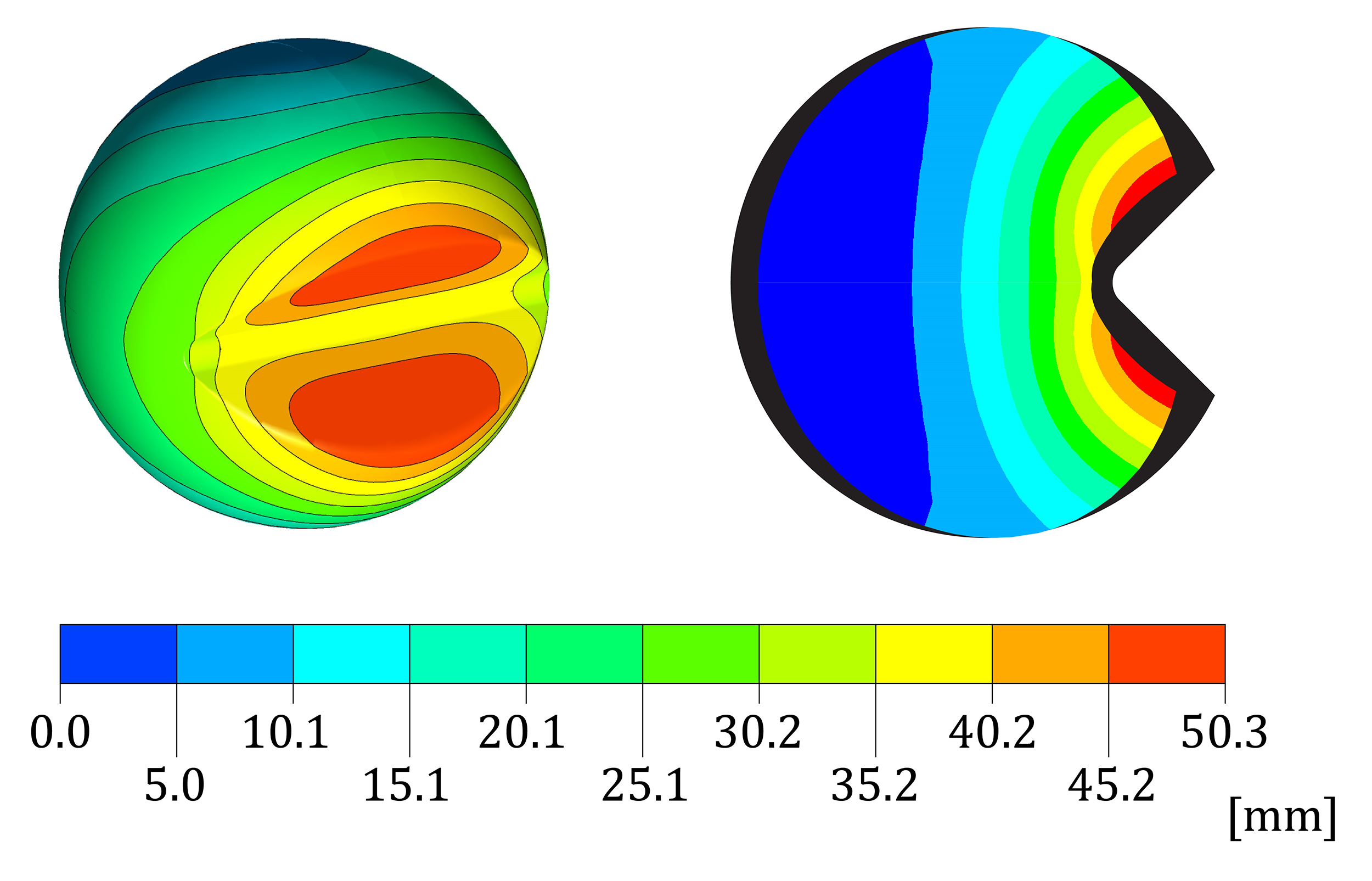}\\
	\includegraphics[width=5.8cm]{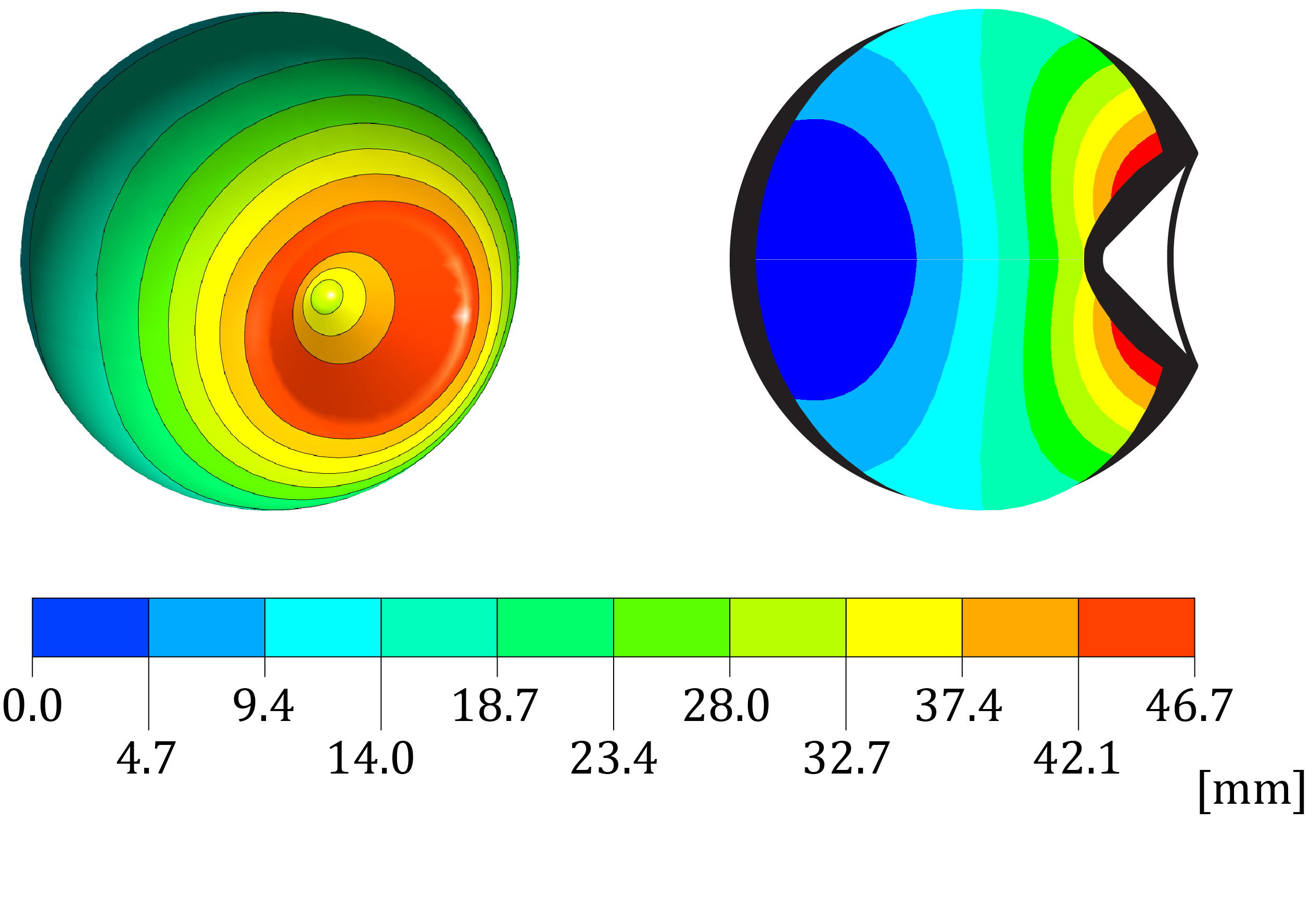}
    \end{tabular}
    \caption{Displacement of points in air flow at the surface of iron SBs with radii R=100 m of three types (a sphere, a sphere with a wedge, and a sphere with conical notch) with same sizes relative to initial positions. Depths of both notches equal  $0.5R$ for the velocity 20~km/s and the air density at 10~km altitude. The initial opening angles are $90^\circ$, the curvature radii of both notches bottoms are $r_0$=5~m. The colour of the gradation scale corresponds to the value of the displacement. Deformed shapes of the SBs with the displacements exaggerated by 500 times to be discernible are plotted on the right images. Initial undeformed 
 shapes are shown schematically by black shadows.}
    \label{fig5}
\end{figure}

The results of calculating the pattern of the SB static deformation in an air flow obtained using the package ANSYS Mechanical are shown in Fig.~\ref{fig5}. We compare deformation of three types of the SB shape: a sphere, a sphere with a wedge-shaped notch, and a sphere with conical recess. As it was mentioned earlier  the last shape is  the result of collision of a small SBs having low gravity  with meteoroid and in this case  the shape of crater is closer to a cone  \citep{Melosh1989,Giese2006,Giese2014} rather than to a spherical recess (shallow bowl). In particular, the  typical example of a small space body with conical craters is Phoebe.

Fig.~\ref{fig5} shows that the SB shape is changed and accompanied by flattening the frontal and rear parts as well as by increase  in the curvature radius of the notch bottom.

We do not present here the dependence of the deformation stress on the opening angle, since it is obvious that the maximum stress is reached at the opening angle of $90^\circ$ which follows from Eq.~\eqref{eq2} 
below. 
In a simplified form, shown in Fig.~\ref{fig1}d, the projection of the aerodynamic force acting perpendicular to the faces of the defect produces  a pressure equal to 
\begin{equation}
    P_\nu=\frac{1}{2}\rho_h V^2 \sin\left(\theta/2\right).
\end{equation}
Here $\rho_h$ is the density of the atmosphere at the altitude $h$, $V$ is the velocity of SB, $\theta$ is the opening angle of the wedge. The sum of the modules of the horizontal projections of this force acting on both sides creates a pressure equal to
\begin{equation} \label{eq2}
    P=\frac{1}{2}\rho_h V^2 \sin\left(\theta\right).
\end{equation}

The stress and deformation calculations for the iron SB were processed with material properties of pure iron: Young's modulus of $2\cdot 10^{11}$~Pa and Poisson's ratio of $0.25$. Inertia relief option was used to compensate the difference between frontal and backward pressure by applying necessary angular and translational accelerations calculated automatically by ANSYS Mechanical to achieve static position for the body in the coordinate system of the simulation. This is common practice for stress calculation for a flying body in a viscous medium. 

\begin{figure}
\centering
    \begin{tabular}{c}
	\includegraphics[width=6cm]{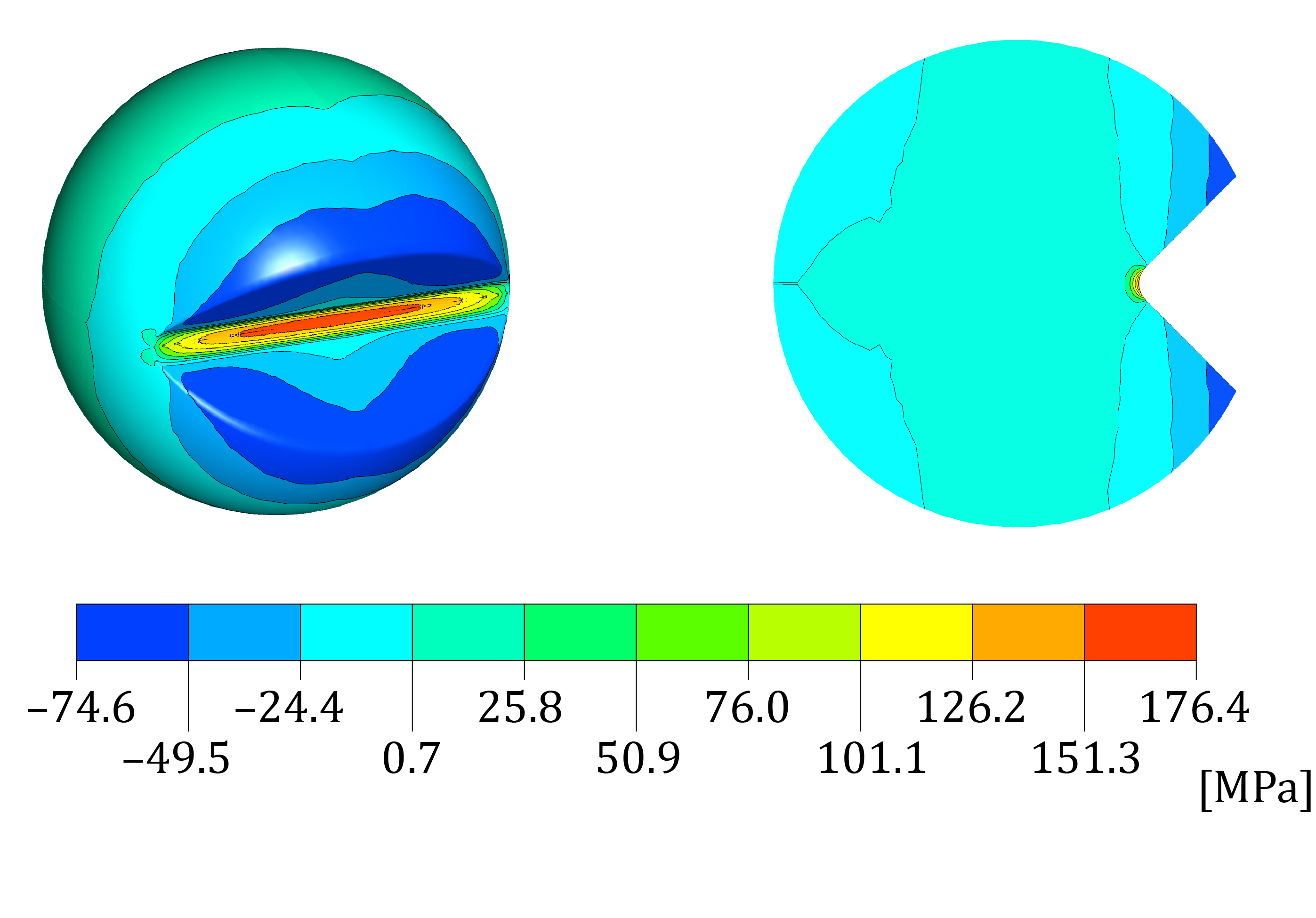}\\
	\includegraphics[width=6cm]{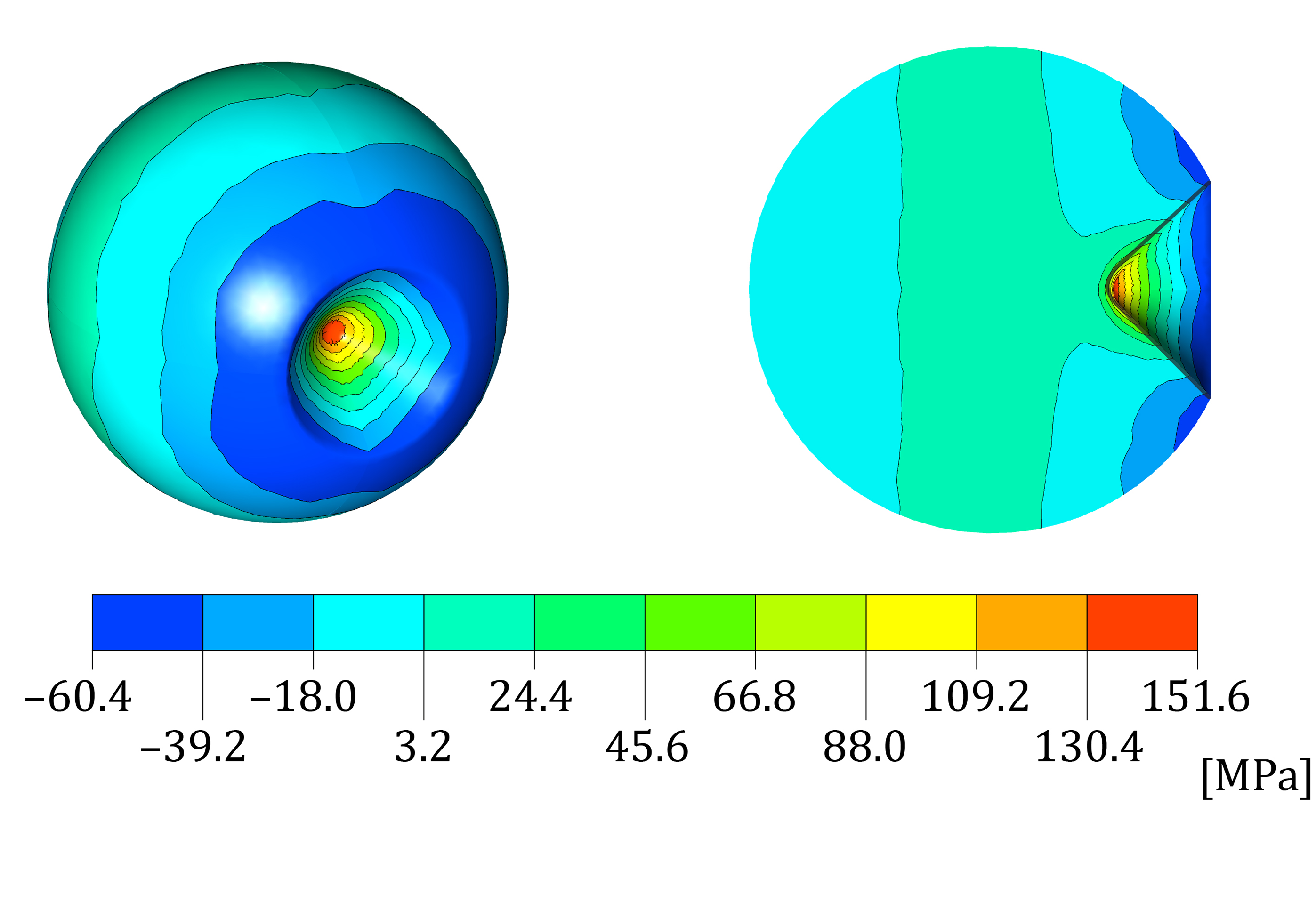}\\
		\end{tabular}
    \caption{Comparison of the first principal stress distributions in the iron SBs  with defects in the form of a wedge and a cone. Here positive values on the gradation scale correspond to a tensile stress, negative values correspond to a compressive stress. Both SBs have $R$=100 m, the depths of the defects are 0.5$R$, the initial opening angles are $90^\circ$, the curvature radii of the notches bottoms are $r_0$=5~m, V=20 km/s.}
    \label{fig6}
\end{figure}

Fig.~\ref{fig6} shows first principal stress for two spherical SBs of the same size with two types of defects -- wedge and cone, which provides maximum tensile stress. This figure clearly shows the appearance of the compression zones around defects and the zones with maximum tensile stress in the bottom part of defects. This pattern indicates a higher resistance to fragmentation of SB with a conical defect compared to the wedge.

\begin{figure}
\centering
	\includegraphics[width=4cm]{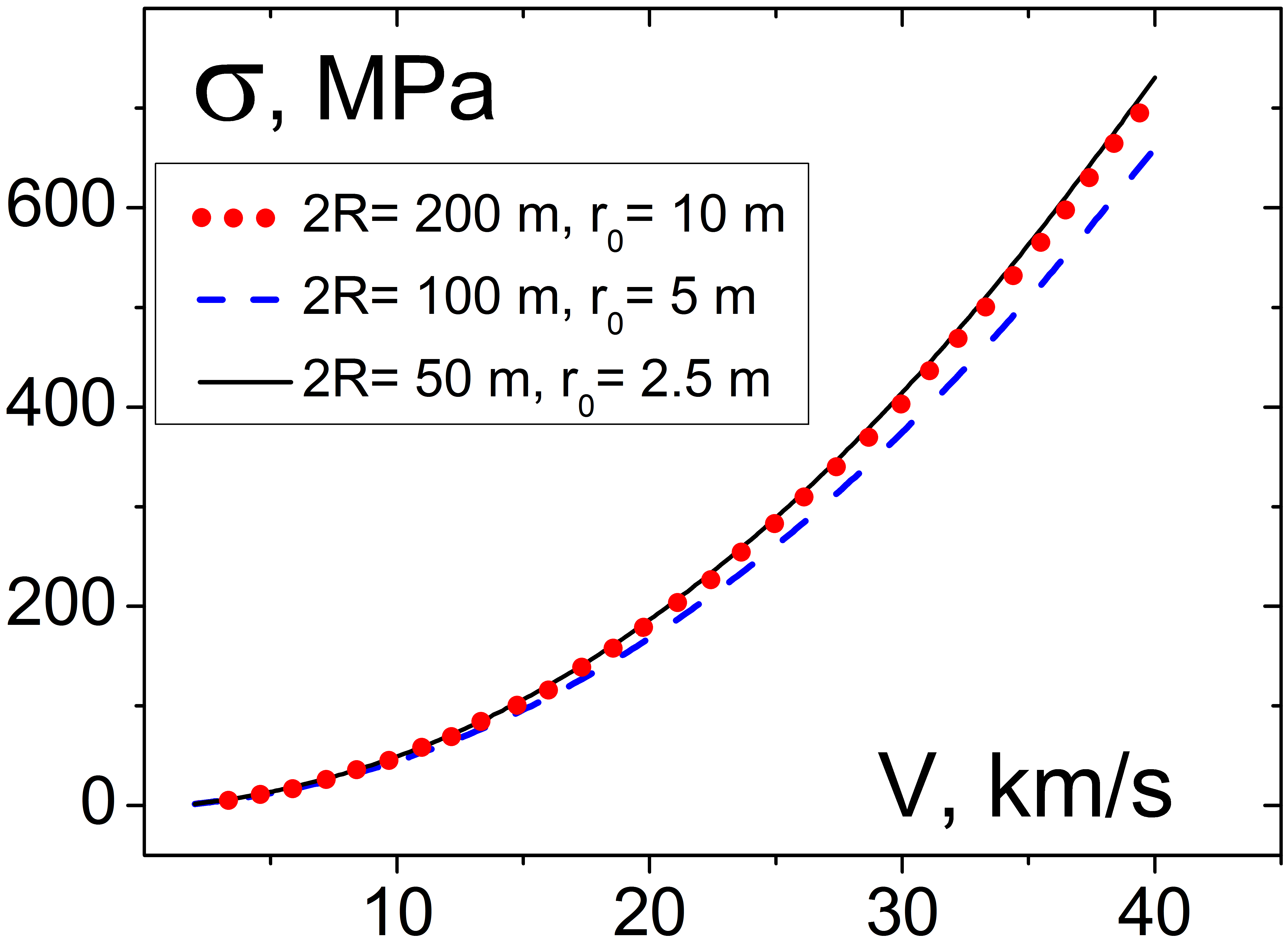}
    \caption{The dependencies of the maximum value of the first principal stress in the iron SBs with wedge-shaped notch and radii $R=100$, 50, and 25~m on the velocity. The curvature radii of the notch bottom $r_0$=10, 5, and 2.5~m. The opening angle is $90^\circ$, the depth of the notch is $h=0.5R$.}
    \label{fig7}
\end{figure}

\begin{figure}
\centering
	\includegraphics[width=4cm]{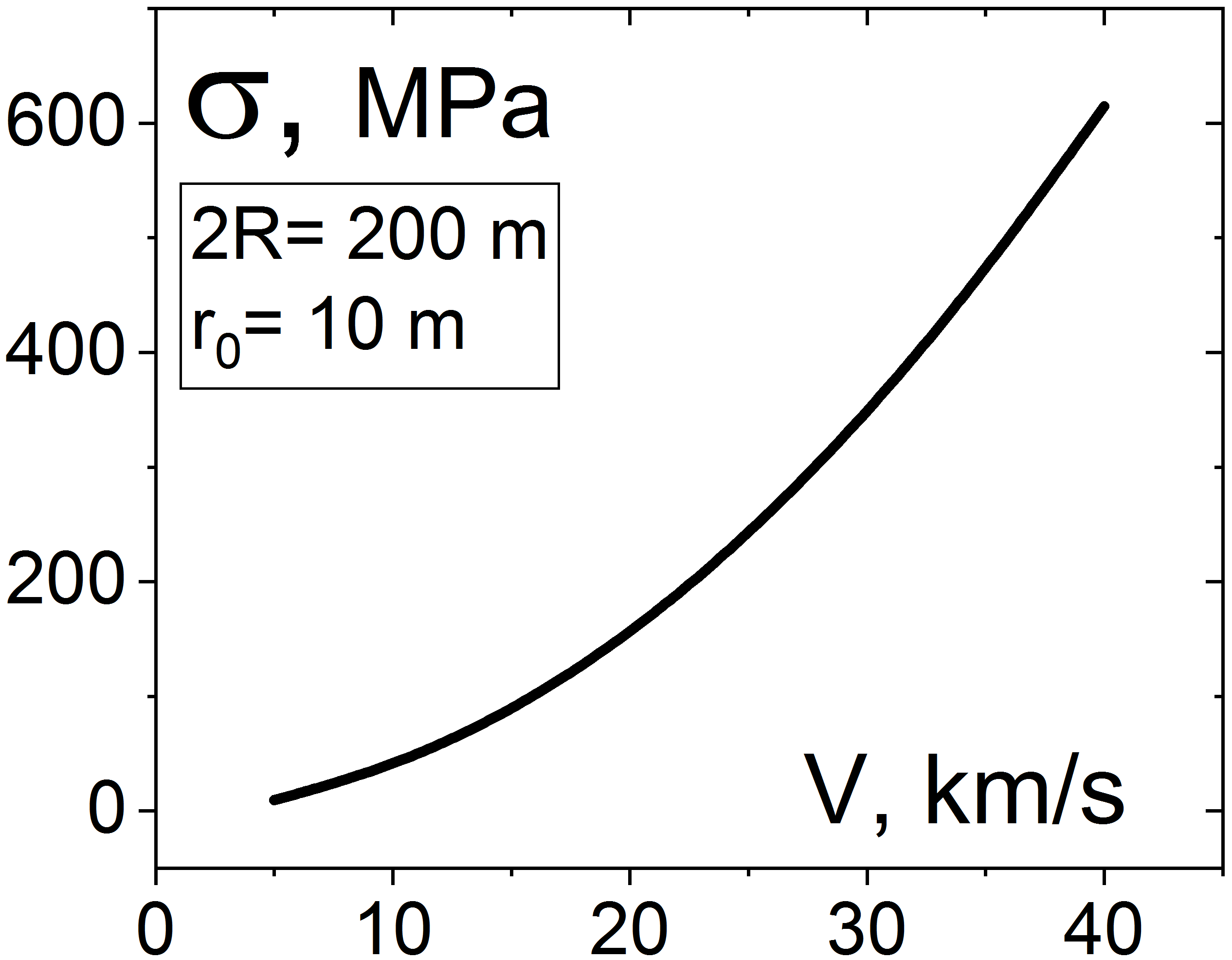}
    \caption{The dependency of the maximum value of the first principal stress in the iron SB with conical notch and radius R=100~m   on the velocity. The curvature radii of the notch bottom $r_0$=10~m. The opening angle is $90^\circ$, the depth of the notch is $h=0.5R$.}
    \label{fig8}
\end{figure}

The effect of velocity on deformation stress in SBs with different surface defects is shown in Figs.~\ref{fig7} and \ref{fig8}. 
Fig.~\ref{fig7} shows the dependence of the maximum value of the deformation stress in the SB sphere with the wedge on the velocity for three values of radius, $R=100$~m (a), 50~m (b), and 25~m (c). Figs.~\ref{fig7} and \ref{fig8} include new parameter $r_0$ -- the curvature radius of the bottom part of the notch (Fig.~\ref{fig9}).  For larger SB sizes the curvature radius  $r_0$ was taken to be large enough to show the range of velocities, when the values of stress are below the tensile stress threshold of destruction. For even larger values of $r_0$ under otherwise the same conditions the integrity of the SB is preserved during the passage through the atmosphere.
Fig.~\ref{fig8} shows the dependence of the maximum value of the deformation stress in the SB sphere with cone and R=100 m on the velocity. 
Comparison of these figures shows minimal difference for equal parameters ($R$, $r_0$, and $V$) -- the stress in SBs with a cone shaped defect is a little smaller than the stress in SBs with a wedge shaped defect.

\subsection{Dependence of the deformation stress value on the curvature radius of the notch bottom}

\begin{figure}
\centering
	\includegraphics[width=4.5cm]{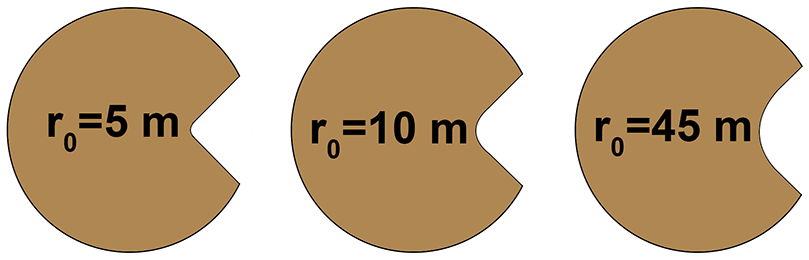}
    \caption{The profiles of  SBs ($R=100$~m) with  wedge-shaped  notch with its depth of $0.5R$, opening angle of $90^\circ$, and different values of the radius of curvature ($r_0$) at the bottom part of the notch: 5~m, 10~m and 45~m.}
    \label{fig9}
\end{figure}

The performed calculations reveal the important feature that affects the value of the deformation stress in SB. The parameter that has a strong effect on the stress magnitude is the radius of curvature ($r_0$) of the bottom part of the notch (Fig.~\ref{fig9}). This parameter plays the same important role as the depth of the notch or its coverage angle.

\begin{figure}
\centering
	\begin{tabular}{cc}
	\includegraphics[width=4cm]{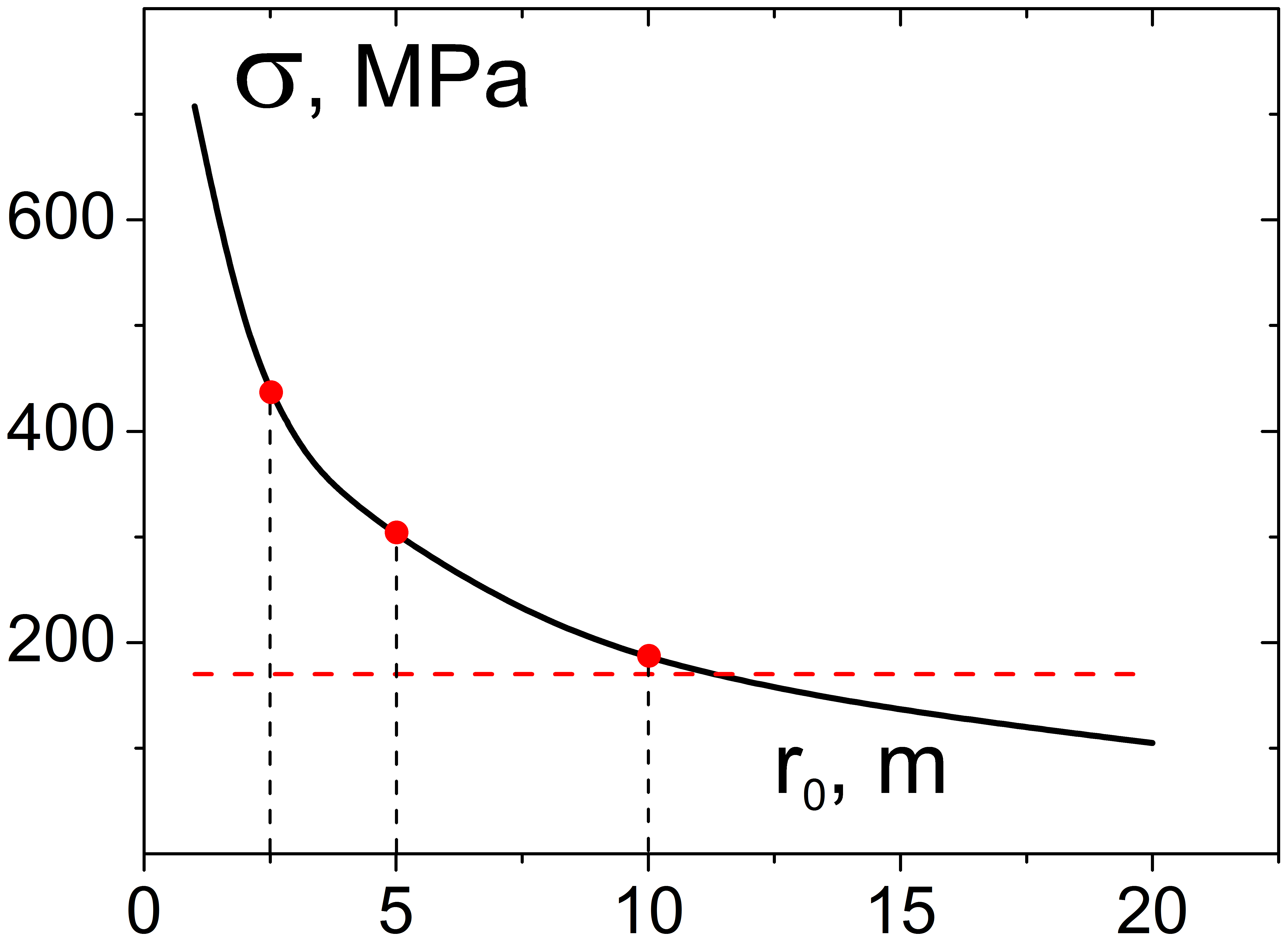}&
	\includegraphics[width=4cm]{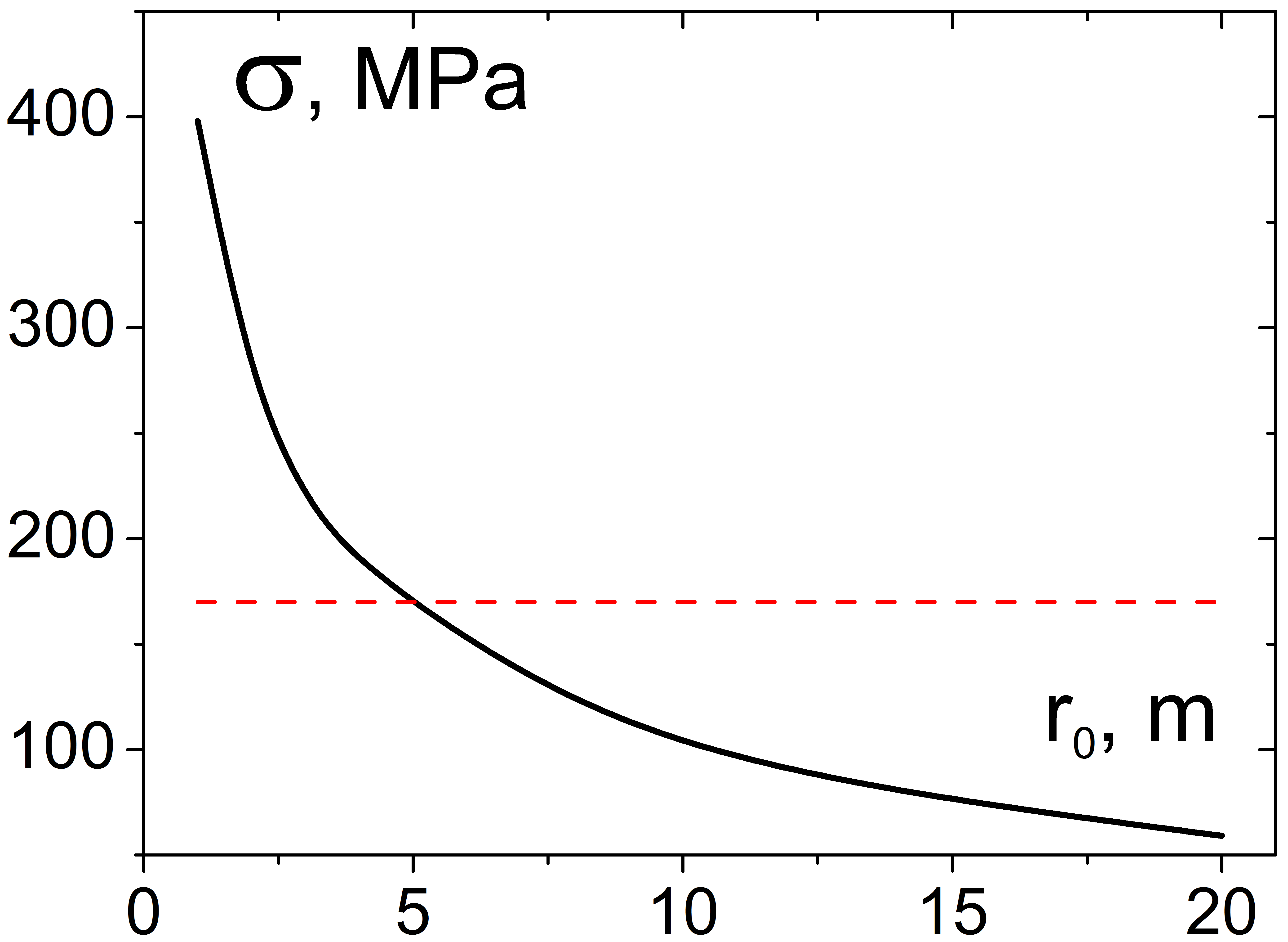}\\
	(a)&(b)\\
	\includegraphics[width=4cm]{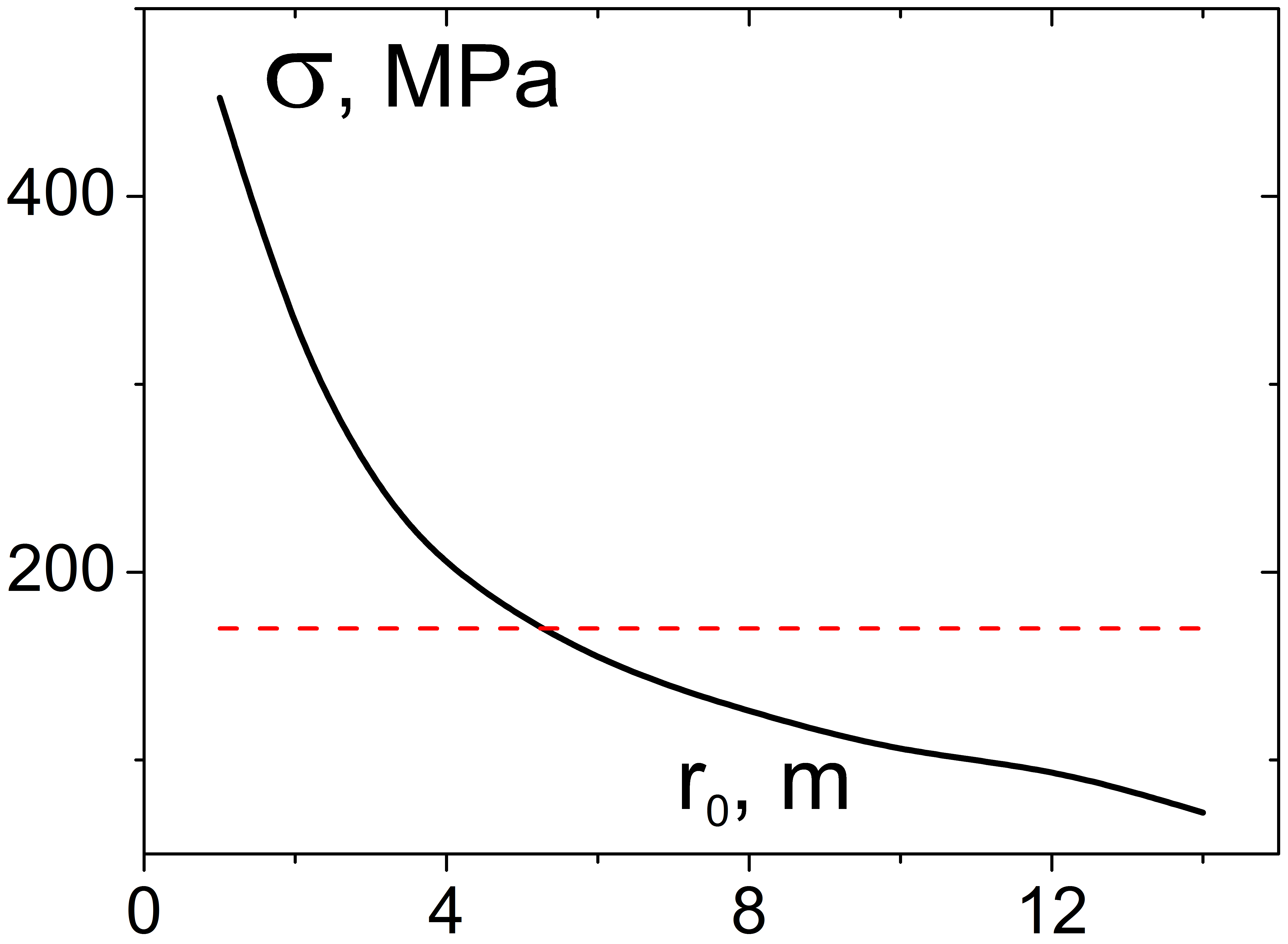}&
	\includegraphics[width=4cm]{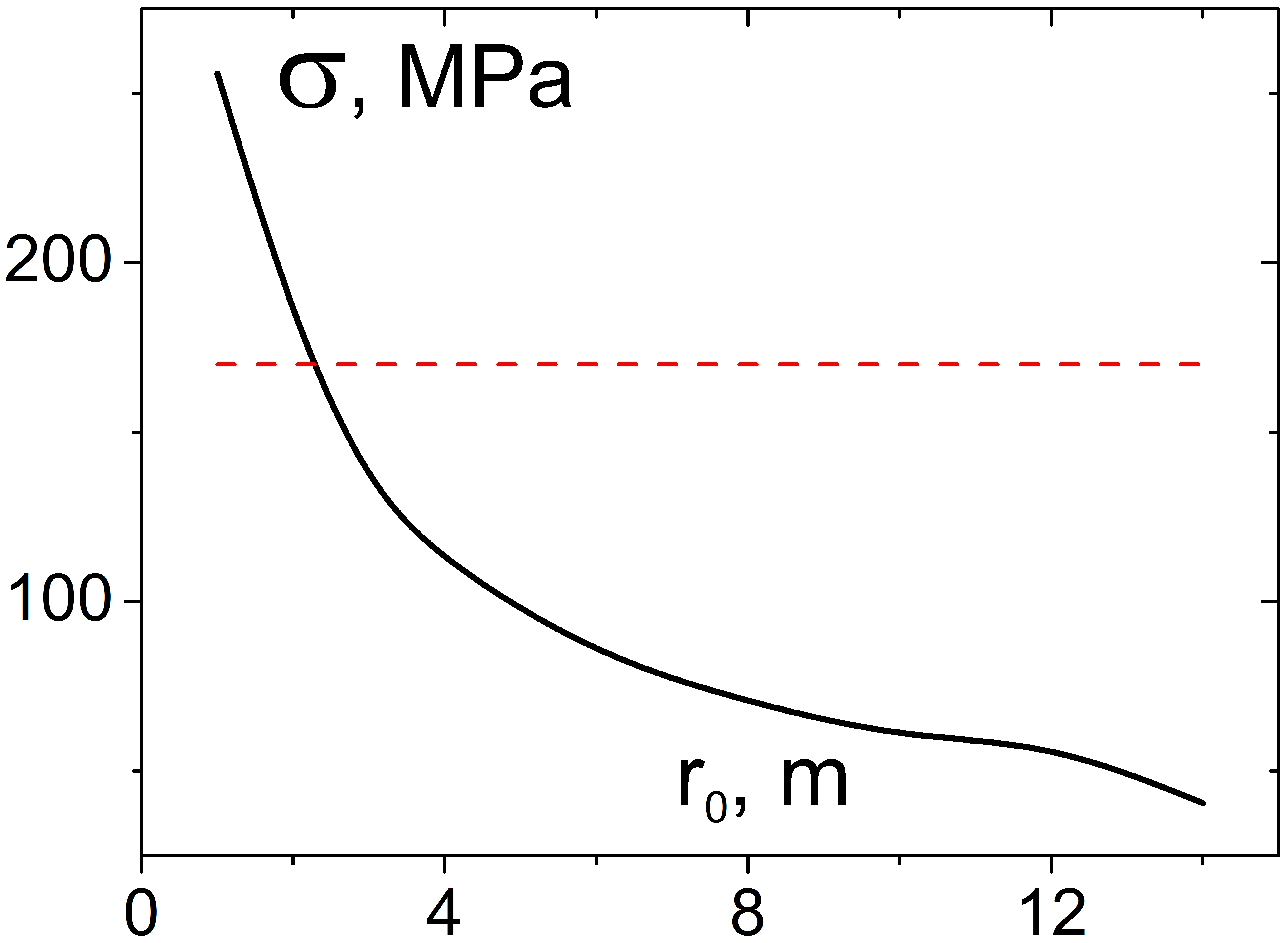}\\
	(c)&(d)\\
	\includegraphics[width=4cm]{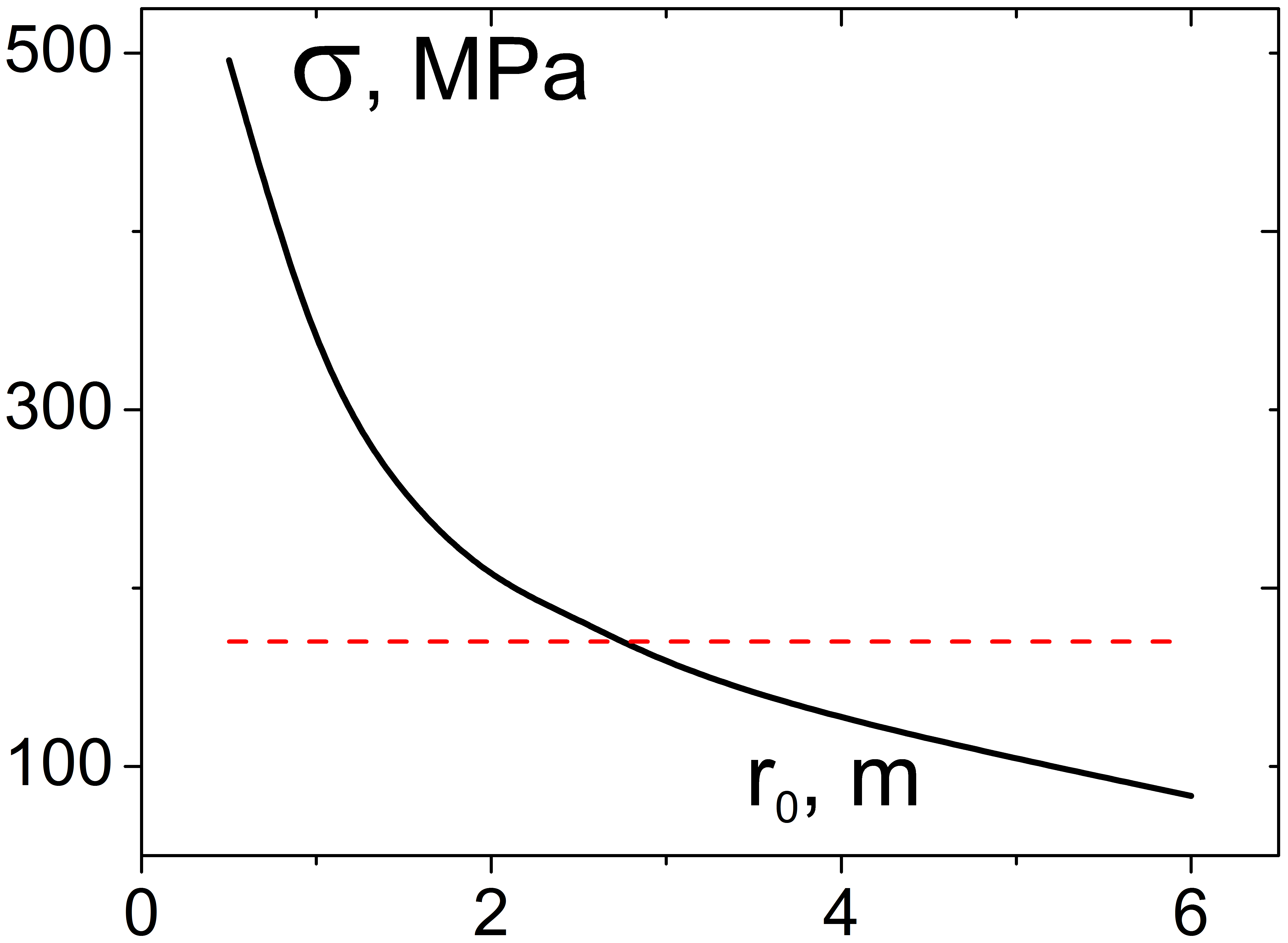}&
	\includegraphics[width=4cm]{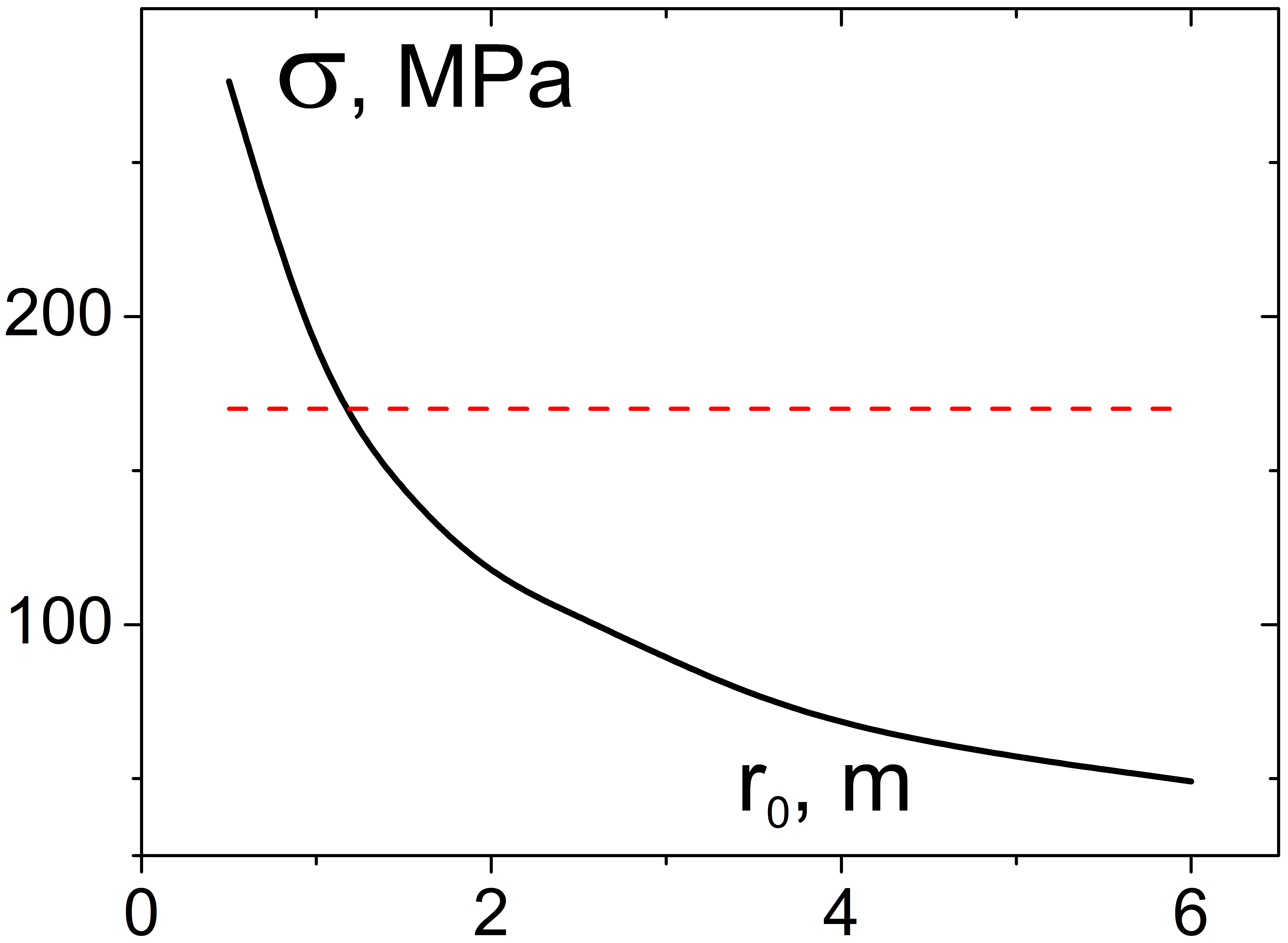}\\
	(e)&(f)
	\end{tabular}
    \caption{Dependencies of the maximum value of the first principal stress in the iron SBs with radii  $R=100$~m (a,b), $R=50$~m (c,d) and $R=25$~m (e,f) on the curvature radius of the bottom part of the wedge-shaped notch ($r_0$) at the SB velocity of 20~km/s (a,c,e) and 15~km/s (b,d,f), the depth of notch $h=0.5R$; the initial opening angle $90^\circ$.}
    \label{fig10}
\end{figure}

Fig.~\ref{fig10} demonstrates 
dependencies of the stress on the curvature radius of the notch for SBs with three radii: $R=100$, 50 and 25~m and for the velocity of SB 20 and 15~km/s. 

Note that this work is a continuation of our recent paper \citet{Khrennikov2019}, which discusses the possibility of through passage of an iron SB across the Earth's atmosphere with preservation of a significant fraction of the initial mass. The dominant mechanism of the mass loss at high velocity in the atmosphere is the high-temperature sublimation of the SB material \citep{Khrennikov2019}. The aim of this work is to estimate the probability of fragmentation of an iron SB during its through passage of the atmosphere with an initial velocity over 15 km/s as an additional mechanism of a mass loss. 
Fig.~\ref{fig10} shows that at velocities V=20 and 15 km/s  significant part of the initial mass can be  preserved after the grazing passage through the atmosphere, and this velocity range is of interest in assessing the possibility of such passage with preservation of the SB integrity. This figure shows that the dependence of stress on the curvature radius is very strong. The smaller $r_0$, the greater the stress for the same external pressure. For smaller SBs calculations were performed for narrower range of curvature radius $r_0$, which resulted in higher stress values in otherwise the same conditions. 

The compressive strength of iron is on average 3--5 times greater than the tensile strength. There is a significant spread of the values for the tensile strength threshold of iron in the literature due to different kinds of iron. Because we aim at evaluating the preservation of the integrity of an SB after through passage across the atmosphere, we choose the most conservative value for the tensile strength threshold of 170~MPa, the minimum one we have found in the literature \citep{Pisarenko1975}. In Fig.~\ref{fig10}, this minimum tensile strength threshold of iron is indicated as a dashed horizontal line. As we can see, the resulting deformation stress can considerably exceed this threshold, and it can be considered as the condition of breaking the integrity of the SB. 

\begin{figure}
\centering
	\includegraphics[width=9cm]{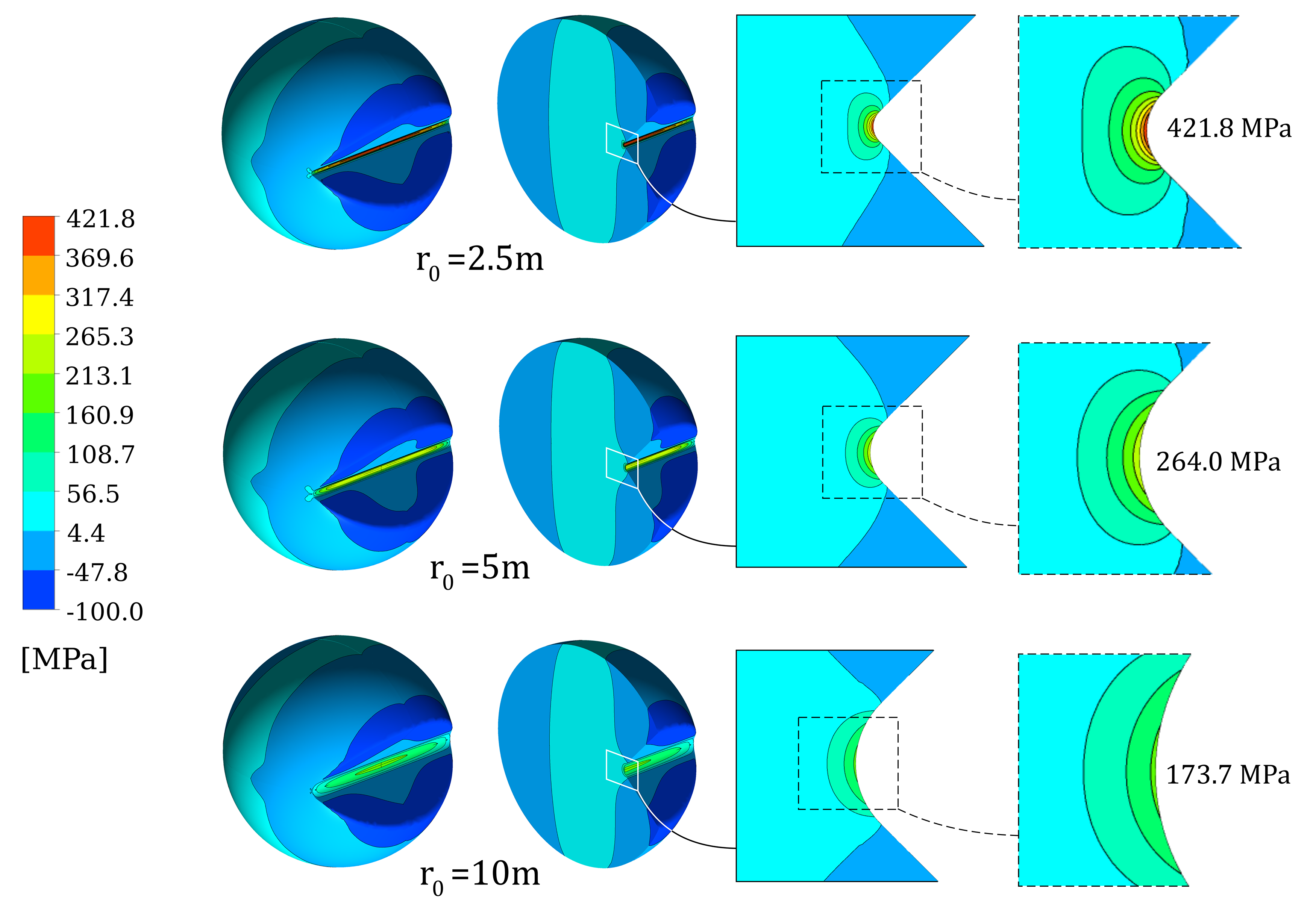}
    \caption{First principal stress in the iron SB with size $2R=200$~m ($h=0.5R$) vs the  curvature radius of the notch bottom ($r_0$) at the velocity 20~km/s and $r_0$=2.5, 5, and 10~m (these points are indicated in Fig.~\protect\ref{fig10}a as red circles).
    Here positive values on the gradation scale correspond to a tensile stress, negative values correspond to a compressive stress.}
    \label{fig11}
\end{figure}

 We can see in Figs.~\ref{fig10} a,c,e that when the  curvature radii of the bottom of the notch exceed values of 3, 7, and 12~m for SBs with the radii of 25, 50, and 100~m, correspondingly, the conditions for fragmentation at $V=20$~km/s are not achieved. 
 
Fig.~\ref{fig11} shows distribution of the deformation stress inside the SB with $2R=200$~m, a wedge-shaped notch depth of $h=0.5 R$, and initial opening angle of the notch $90^\circ$, at three values of the curvature radius $r_0$ and distribution of the stress near the bottom. As we can see, the maximum stress arises  near the bottom of the notch. The maximum stress can significantly exceed the rupture tensile threshold value. This localization of the stress leads to tearing the bottom of the notch in the SB and subsequent redistribution of the stress into deeper, underlying layers of the SB, which will be opened after breaking the upper layer as the SB progresses along the trajectory.

\begin{figure}
    \centering
    \includegraphics[width=5cm]{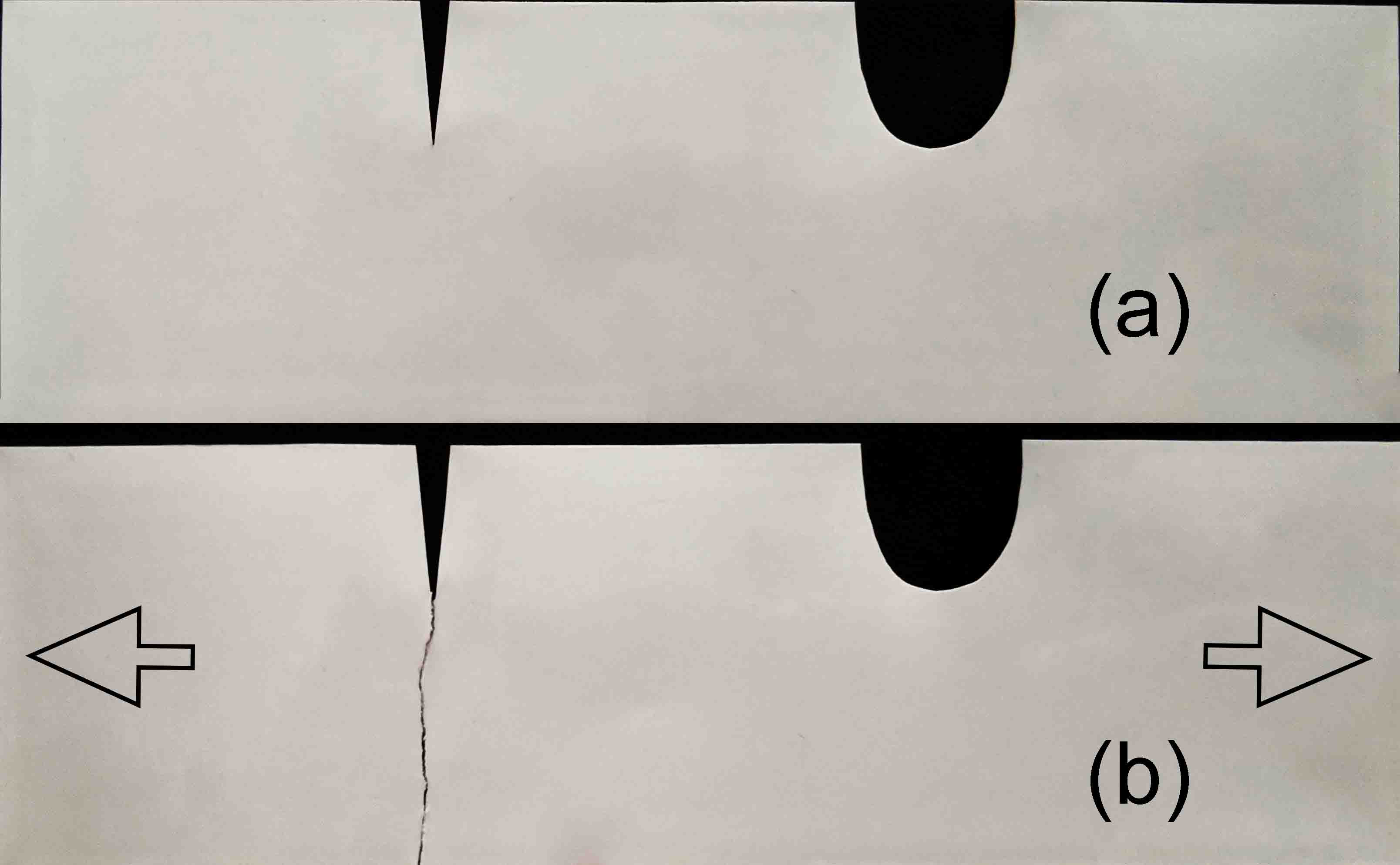}
   \caption{Fragment of a sticking tape with two types of incisions  of same depth with small and large  curvature radii in bottom part: (a) -- before stretching, (b) -- after stretching the tape with the break position (arrows show applied forces.}
  \label{fig12}
\end{figure}

Qualitatively, the effect of the curvature radius $r_0$ on fragmentation can be explained by the similarity with the effect of the small narrow transverse incision made in the stretched tape of  arbitrary material. 
Fig.~\ref{fig12} shows how this incision  dramatically reduces the threshold force needed to break up the fragment of sticking tape whereas the wider incision with larger curvature radius does not change the threshold significantly.

\subsection{Dependence of the  deformation stress on the depth of a notch}
\begin{figure}
\centering
	\begin{tabular}{c}
	\includegraphics[width=4cm]{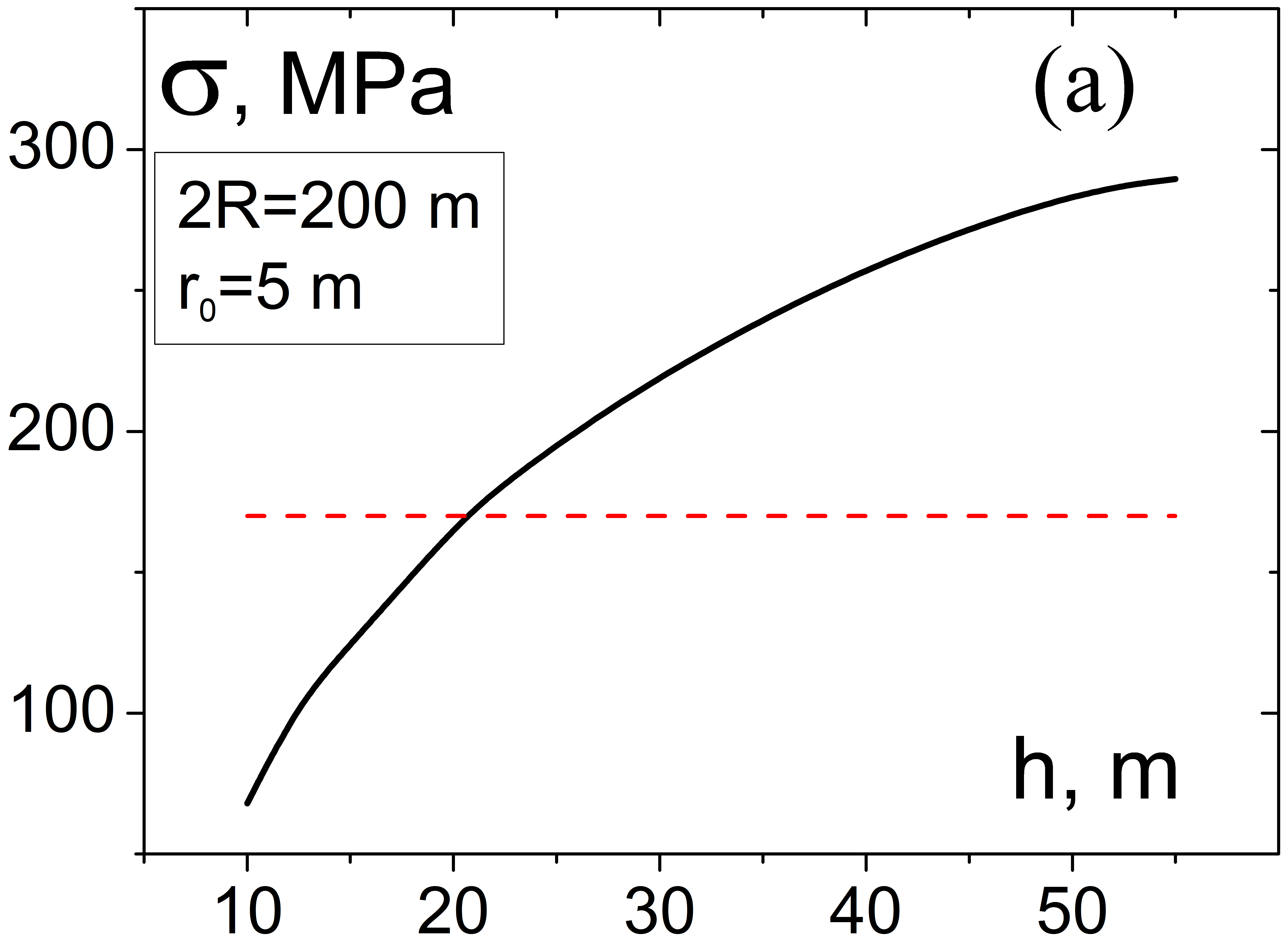}\\
	\includegraphics[width=4cm]{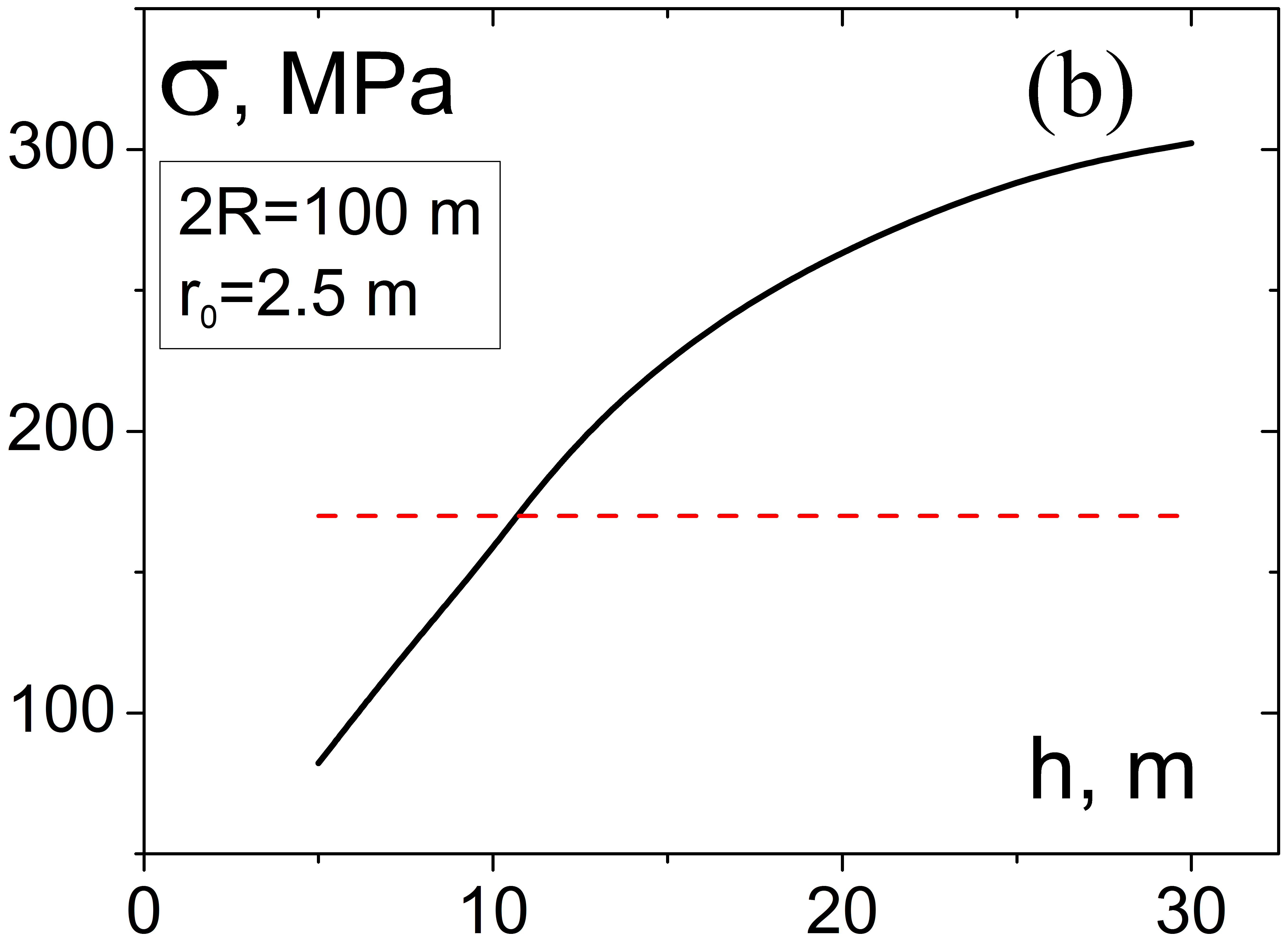}\\
	\includegraphics[width=4cm]{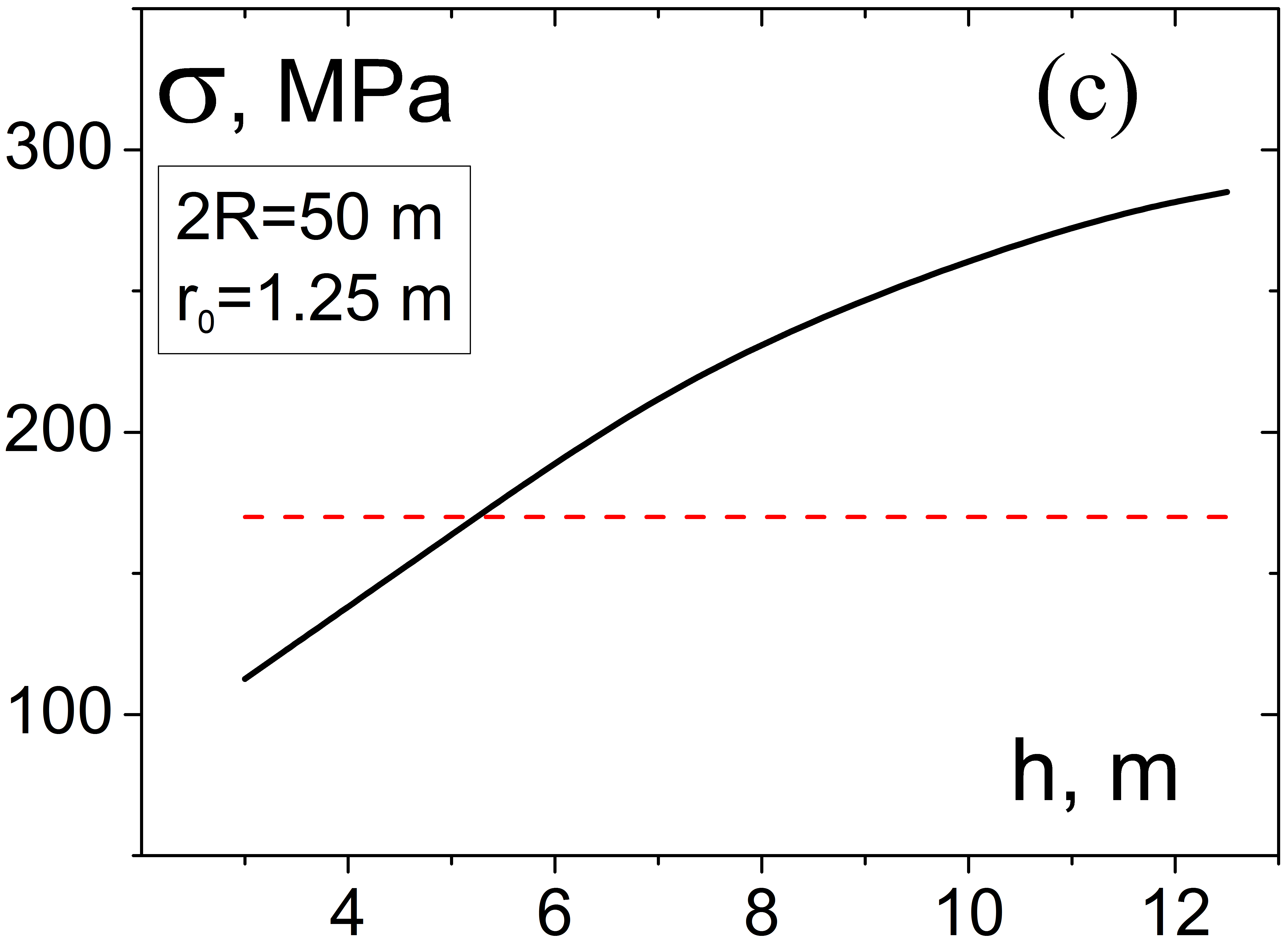}
	\end{tabular}
    \caption{The dependencies of the maximum value of the first principal stress in the iron SBs with radii $R=100$~m (a), 50~m (b), and 25~m (c) on the depth of the notch at the velocity 20~km/s and $r_0=5$~m (a), $r_0=2.5$~m (b) and $r_0=1.25$~m (c).}
    \label{fig13}
\end{figure}

Fig.~\ref{fig13} shows the dependence of the maximum value of the deformation stress in the SB on the depth of the notch $h$ and, respectively, on the area of the side surfaces of the notch. The calculations were made for SBs with radii $R=100$, 50 and 25~m for the velocity of $20\,\mbox{km/s}$ with an optimum opening angle of $90^\circ$. The depth of the notch is varied from $h=0.1R$ to $0.5R$. We can see that the smaller value of the curvature radius of the notch bottom compensates the effect of increasing the notch depth on the amplitude of the strain stress in SBs with different sizes.
Fig.~\ref{fig13} demonstrates  that despite  significantly higher aerodynamic pressure on the surface of larger SBs and greater deformation stress at the same value of $r_0$, the decrease in $r_0$ in smaller SBs makes it possible to equalize maximum  stress in SBs of different sizes and to  increase the probability of their fragmentation.   

\subsection{Deformation of stone space bodies with different surface shapes} 

\begin{figure}
\centering
    \begin{tabular}{c}
	\includegraphics[width=6cm]{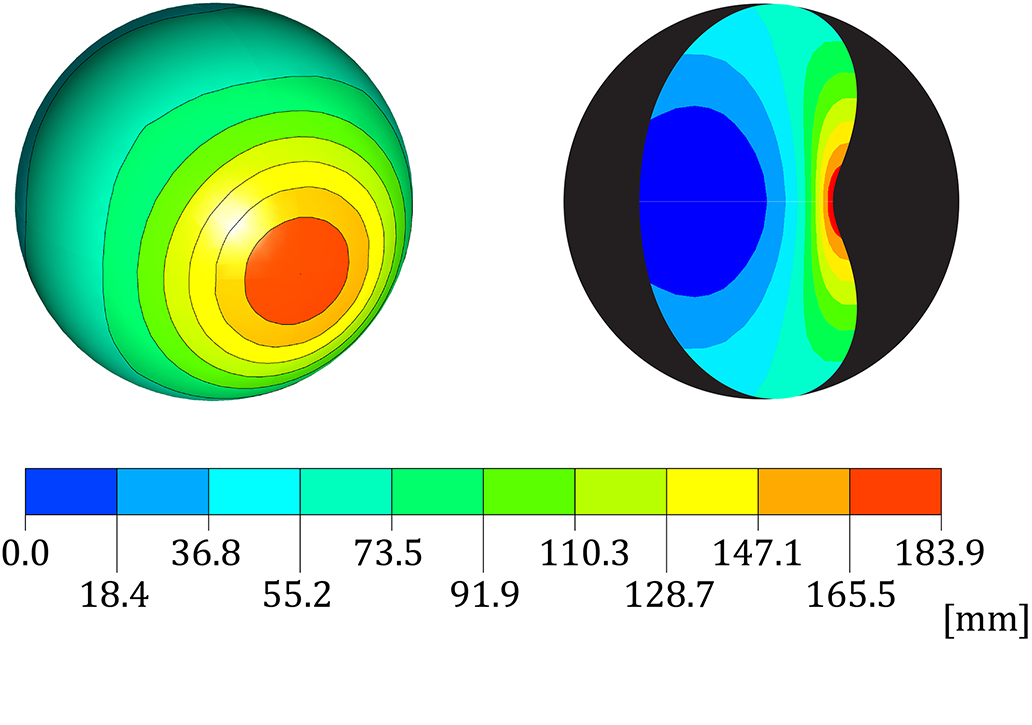}\\
	\includegraphics[width=6cm]{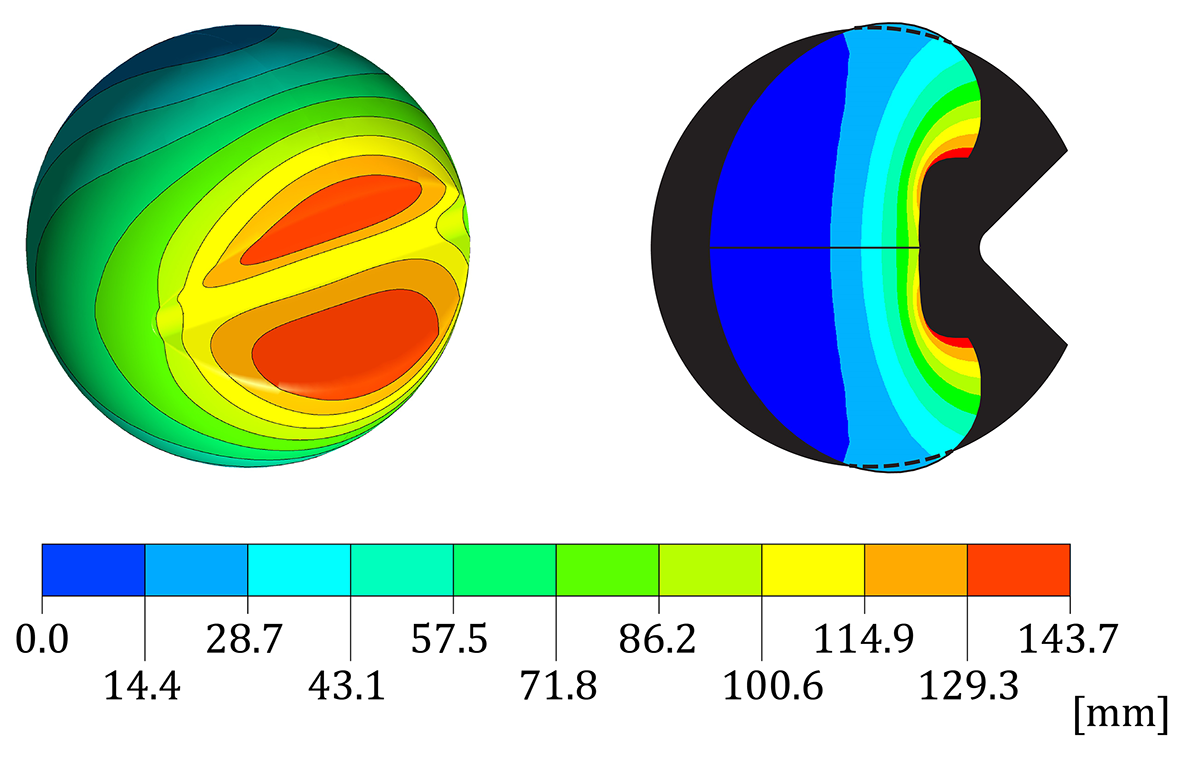}\\
	\includegraphics[width=6cm]{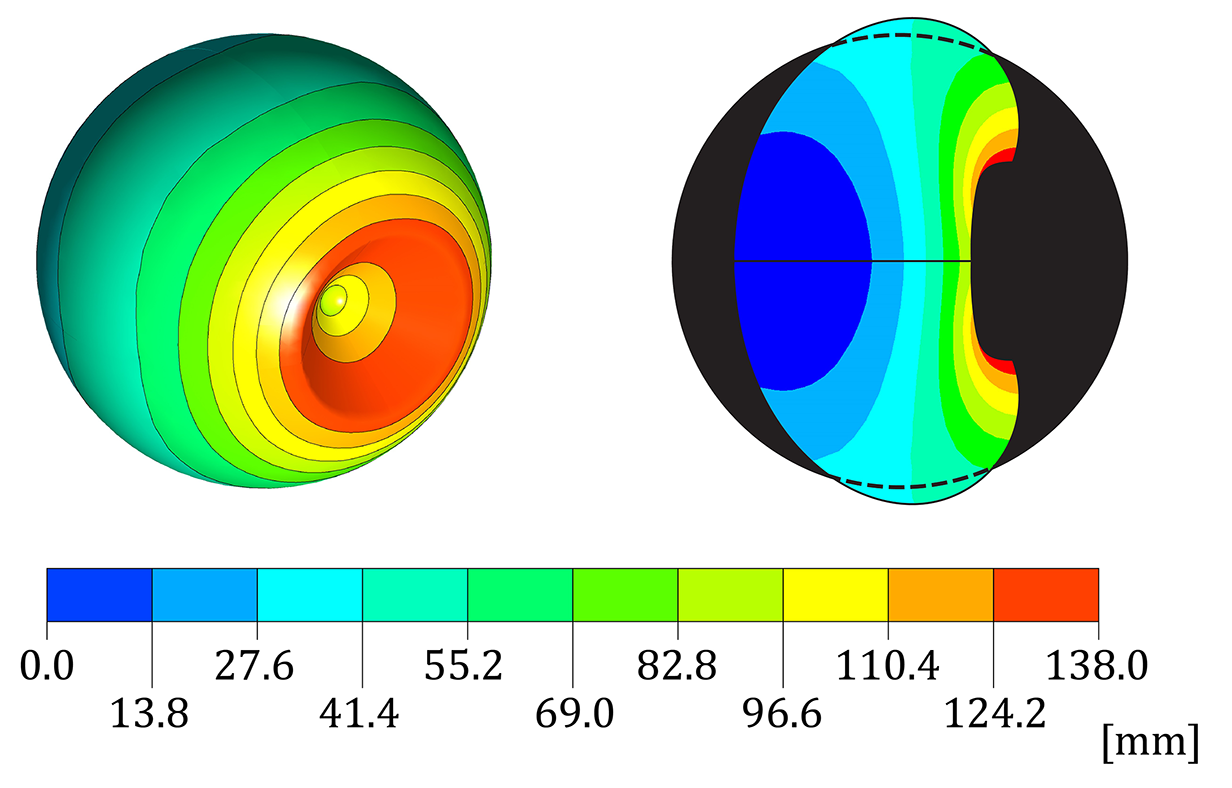}
	\end{tabular}
    \caption{Displacement of points at the surface of the granite SBs with radii $R$=100~m relative to initial positions for the depths of both notches equal to $0.5R$, for the velocity 20~km/s and for the air density at 10~km altitude. The initial opening angles are $90^\circ$, the curvature radii of the notch bottoms are $r_0=5\,\mbox{m}$. The colour of the gradation scale corresponds to the value of displacement.
Deformed shapes of the SBs with the displacements exaggerated by 500 times to be discernible are plotted on the right panels. Initial undeformed 
 shapes are shown schematically by black shadows.}
    \label{fig14}
\end{figure}

Fig.~\ref{fig14} demonstrates deformation of stone SBs of the same size with different shapes: sphere, sphere with wedge-shaped notch, and sphere with conical notch. Calculations for stone SBs were carried out with material properties of granite  -- Young's modulus  equals $7\cdot10^{10}$~Pa and Poisson's ratio of $0.25$. In comparison with iron SBs~(Fig.~\ref{fig5}), the deformation of the stone SBs of the same size and shape is three times higher.  Note that the first principal stress distributions in both iron and stone SBs are equal and are determined by aerodynamic flow speed. Therefore, one can use the dependencies of the maximum stress shown in Figs.~\ref{fig7}, \ref{fig8}, \ref{fig10}, and \ref{fig13} for stone SBs, only to replace the value of the stress threshold for disruption (horizontal dashed line in Figs.~\ref{fig10} and~\ref{fig13}) by the appropriate value for granite or other kind of stone. The difference in materials (iron and stone)  affects only the value of deformation which in stone bodies is three times higher and depends on elastic properties of materials. 

In our final comments, which can be considered as the next steps of study, we can mention  the  problems  of deformation and  stress of stone space bodies taking into consideration the factor of internal pressure produced by penetration  of ultrahigh pressure gas from the shock wave boundary layer into  micro-cracks. Besides that, one can study variations of the deformation stress with the position of the recess on the frontal surface of SB and angle relative to the air flow. 

An important step in future investigations  would be the  simulation of  the fragmentation process in real-time by ANSYS explicit dynamics tools.

\section*{Summary}

The methods employed in our paper made it possible to determine the conditions for fragmentation of the SB or for preservation of its integrity when passing through the planetary atmosphere in various situations. The results obtained support the following statements: 
\begin{enumerate}[label={\arabic*.}]
    \item The transverse aerodynamic forces causing destruction of the iron SB and arising on the defects of its frontal surface in the form of  notch are able to overcome the tensile strength threshold of iron. In order to cause the destruction of  the SB, these defects should be in the form of the extended surface wedge or cone with the depth over 1/2 of the SB mean radius or larger even at the velocity over 20--25 km/s. Of all cases, the most susceptible for fragmentation of the SB are when the opening angle of a wedge or a cone equals $90^\circ$. 
  The fragmentation conditions for the model defects in the shape of a spherical sector or of a cone are becoming more favourable  with the increase in depth of the notch and, correspondingly, in the area of the  aerodynamic surfaces, as well as with the increase in the initial velocity of the entry into the atmosphere. 
    \item The curvature of the bottom of the surface notch of SB is an important geometric parameter that determines the conditions for fragmentation. The smaller this value, the higher the strain stress in the bottom of the notch and the more likely the SB to fragment.
    \item Upon the entry of the SB into the denser layers of the atmosphere, the notches in the SB can be eroded because of the sublimation (ablation) of the material, which is accompanied by a gradual decrease of aerodynamic forces acting to disintegrate the SB. However, due to the significant initial mass and size, the substantial parts of the wedge can  be preserved.  
    \item The obtained dependencies make it possible to estimate the probability of fragmentation of SBs consisting of materials with the tensile strength threshold  lower than for iron~--- 50~MPa or less. 
    \item The conditions for preserving the integrity of a large iron SB during its passage through  the atmosphere at a minimum altitude of 10---15 km are as follows: initial velocities within  15~---20 km/s and the absence of deep  irregularities (over 0.2--0.3 of the SB radius) of the surface relief. However taking into account the wide range of the available values of the threshold strength of meteorite  materials exceeding 300--400 MPa~\citep{Svetsov1995}, the integrity of some of the large iron space bodies can be preserved at higher velocities as well.
   \item  As we have calculated in our work, the radius of this iron asteroid body could have been from 50 to 100~m to avoid disintegration in the atmosphere and to preserve significant part of the initial mass. The most important conclusion of this work is that we have confirmed the realism of the hypothesis that the Tunguska phenomenon can be likely  associated with a grazing passage of an iron asteroid body across the Earth's atmosphere with velocities below the fragmentation threshold~-- 20~km/s at the minimum altitude of 10--15~km and its subsequent escape into outer space to the orbit around the Sun~\citep{Khrennikov2016,Khrennikov2019}. 
        
\end{enumerate}

\section*{Acknowledgements}

The article benefited from many suggestions and comments made in the constructive report by the reviewer, Dr. Darrel Robertson, whom we thank for careful reading of the manuscript.
We are grateful to Doug Black of Hamilton, Canada for correcting English in the final version of the manuscript.




\bibliographystyle{mnras}

\bibliography{biblio}




\bsp	
\label{lastpage}
\end{document}